\documentclass[useAMS,usenatbib]{mnras}

\usepackage{graphicx}
\usepackage{float}
\usepackage{hyperref}
\usepackage{color}
\usepackage{amsmath}
\usepackage{amssymb}
\usepackage{mathrsfs}
\usepackage{epsfig}
\usepackage{epstopdf}
\usepackage{multirow}
\usepackage{threeparttable}

\newcommand{\Lx}{$L_X$}
\newcommand{\Lagn}{$L_{\rm AGN}$}
\newcommand{\Ledd}{$\lambda_{\rm Edd}$}
\newcommand{\z}{$z$}
\newcommand{\Mstar}{$M_\ast$}
\newcommand{\Msun}{$M_\odot$}
\newcommand{\ergps}{erg s$^{-1}$}

\title[A mass-dependent AGN Eddington ratio distribution]{Evidence for a mass-dependent AGN Eddington ratio distribution via the flat relationship between SFR and AGN luminosity}

\author[E.~Bernhard et al.]
{\parbox{\textwidth}{E.~Bernhard$^{1}$,\thanks{E-mail: \texttt{ebernhard1@sheffield.ac.uk}}
    J.R.~Mullaney$^{1}$, J.~Aird$^{2}$, R.C.~Hickox$^{3}$, M.L.~Jones$^{3}$, F.Stanley$^{4}$, L.P.~Grimmett$^{1}$, and E.~Daddi$^{5}$}
\vspace{0.4cm}\\\
\parbox{\textwidth}{
  $^{1}$Department of Physics $\&$ Astronomy, University of Sheffield, Sheffield S3 7RH, UK\\
  $^{2}$Institute of Astronomy, University of Cambridge, Madingley Road, Cambridge, CB3 0HA\\
  $^{3}$Department of Physics and Astronomy, Dartmouth College, Hanover, NH 03755, USA\\
  $^{4}$Department of Space Earth and Environment, Chalmers University of Technology, Onsala Space Observatory, SE-43992 Onsala, Sweden\\
  $^{5}$CEA Saclay, Laboratoire AIM-CNRS-Universit\'{e} Paris Diderot, Irfu/SAp, Orme des Merisiers, 91191 Gif-sur-Yvette, France\\}}

\begin{document}

\date{Accepted XXX; Received XXX; in original form XXX}

\maketitle

\begin{abstract}
The lack of a strong correlation between AGN X-ray luminosity (\Lx; a proxy for AGN power) and the star formation rate (SFR) of their host galaxies has recently been attributed to stochastic AGN variability. Studies using population synthesis models have incorporated this by assuming a broad, universal (i.e. does not depend on the host galaxy properties) probability distribution for AGN specific X-ray luminosities (i.e. the ratio of \Lx\ to host stellar mass; a common proxy for Eddington ratio). However, recent studies have demonstrated that this universal Eddington ratio distribution fails to reproduce the observed X-ray luminosity functions beyond \z$\sim$1.2. Furthermore, empirical studies have recently shown that the Eddington ratio distribution may instead depend upon host galaxy properties, such as SFR and/or stellar mass. To investigate this further we develop a population synthesis model in which the Eddington ratio distribution is different for star-forming and quiescent host galaxies. We show that, although this model is able to reproduce the observed X-ray luminosity functions out to \z$\sim$2, it fails to simultaneously reproduce the observed flat relationship between SFR and X-ray luminosity. We can solve this, however, by incorporating a mass dependency in the AGN Eddington ratio distribution for star-forming host galaxies. Overall, our models indicate that a relative suppression of low Eddington ratios (\Ledd$\lesssim$0.1) in lower mass galaxies (\Mstar$\lesssim 10^{10\--11}$~\Msun) is required to reproduce both the observed X-ray luminosity functions and the observed flat SFR/X-ray relationship.

\end{abstract}


\section{Introduction}
\label{sec:introduction}

Most galaxies host a central super massive black hole (hereafter SMBH), the masses of which display a tight correlation with their host bulge masses, implying a co-evolution between SMBHs and their host galaxies \citep[e.g.][]{Kormendy1995, Magorrian1998, Merritt2000,Kormendy2011}. Furthermore, the similar redshift evolution of the total SMBH accretion rate density and the total star-formation rate density strongly suggests the existence of a link between SMBH accretion and star-formation at galactic scales \citep[e.g.][]{Heckman2004, Merloni2004, Silverman2008a, Aird2010, Assef2011}. However, the precise connection between the two processes remains poorly understood.

One means of connecting SMBH and galaxy growth that has gained popular support over the past two decades is for accreting SMBHs (observed as active galactic nuclei, AGNs) to directly influence the star formation rates of their hosts (hereafter SFRs) via a variety of feedback mechanisms \citep[see the review of][for details on the feedback mechanisms]{Fabian2012}. However, a major difficulty in identifying the precise role that SMBH accretion has on influencing SFR is the stochastic nature of AGNs \citep[e.g.][see also \S\,4.3 of \citealt{Stanley2015}]{Aird2013, Hickox2014}. It is argued that this randomness is the reason why most studies that have explored the relationship between SFR and X-ray luminosity (a proxy for SMBH accretion power) have reported no evidence of a strong correlation between these parameters, at least for moderate luminosity AGNs which form the majority of the population \citep[i.e. $10^{42}<L_{X}<10^{45}~{\rm erg~s}^{-1}$; e.g.][]{Lutz2010, Harrison2012, Mullaney2012a, Rosario2012, Rovilos2012, Santini2012, Azadi2015, Stanley2015}. Recently, using optically selected AGNs, \cite{Stanley2017} reported an enhancement of SFR among the highest luminosity AGNs (i.e. \Lx$> 10^{45}$~\ergps). However, they demonstrated that this is a direct consequence of the most luminous AGNs residing in more massive host galaxies, meaning the enhanced SFR is a consequence of the relationship between stellar mass and SFR \citep[hereafter referred to as the main sequence, MS; e.g.][]{Daddi2007, Rodighiero2011, Schreiber2015}.

A major problem in investigating the relationship between AGN and SFR is that observations are often subjected to biases (e.g. flux limited samples, rarity of high accretion rate AGNs). Therefore, many studies investigating the connection between SMBH and galaxy growth have adopted a modelling approach \citep[e.g.][]{Aird2013, Conroy2013, Veale2014, Caplar2015, Jones2017, Weigel2017}. In these models, the stochastic nature of AGNs is often implemented via a broad specific AGN X-ray luminosity (i.e. the X-ray luminosity relative to the host stellar mass; \Lx/\Mstar) distribution that, once combined with a mass function, reproduces the observed X-ray luminosity function of AGNs. Overall, these studies find that below \z$\sim$1 the specific X-ray luminosity distribution is well-represented by a broad, universal (i.e. the same distribution, irrespective of SFR or stellar mass) distribution, described by a broken power-law (or similar; e.g. a \citealt{Schechter1976} function), the normalisation of which increases with increasing redshifts \citep[e.g.][]{Aird2013,Veale2014,Jones2017}. These models are also often successful at reproducing the observed flat relationship between SFR and X-ray luminosity \citep[][]{Hickox2014,Veale2014,Stanley2015}. However, \cite{Jones2017} recently found that using a universal broad distribution for the specific X-ray luminosity distribution of AGNs cannot reproduce the X-ray luminosity functions of \cite{Aird2010} beyond \z$\sim$1.2. They argue for the need of a more complicated specific X-ray luminosity distribution of AGNs beyond that redshift. Furthermore, recent studies have reported that the specific X-ray luminosity distribution for AGNs in star-forming hosts significantly differs from that of AGNs in quiescent hosts \citep[e.g.][]{Georgakakis2014, Aird2017, Wang2016}. Overall, they report a higher contribution from star-forming AGN hosts to the total specific X-ray luminosity distribution out to \z$\sim$2.

Motivated by the above findings, we explore a model for the specific X-ray luminosity distribution split between star-forming and quiescent galaxies. After demonstrating that this simple model cannot simultaneously reproduce both the X-ray luminosity functions {\it and} the flat relationship between SFR and X-ray luminosity out to \z$\sim$2, we introduce a mass dependency for the specific X-ray luminosity distribution of star-forming AGN hosts, as recently observed \citep[e.g.][]{Bongiorno2016, Aird2017, Georgakakis2017}. Hereafter, we refer to our model specific X-ray luminosity distribution split between star-forming and quiescent galaxies as ``mass-independent'', and to our model that includes the mass dependency for the specific X-ray luminosity distribution of AGNs in star-forming galaxies as ``mass-dependent''.

In \S\,\ref{sec:method} we describe our method to infer the specific X-ray luminosity distribution for both the mass-independent and the mass-dependent model. Subsequently, in the same section, we present how we incorporate host galaxy properties. We then report the results for the mass-independent model in \S\,\ref{subsec:mass_indep_results}, and for the mass-dependent model in \S\,\ref{subsec:mass_dep_results}. We discuss the general implications of our results in \S\,\ref{sec:discussion}, and conclude in \S\,\ref{sec:conclusion}. Throughout, we assume the specific X-ray luminosity (i.e. \Lx/\Mstar) to be proportional to Eddington ratio (i.e. the ratio between bolometric AGN luminosity and Eddington luminosity, \Lagn/$L_{\rm Edd}$), converting the X-ray 2-10 keV luminosity into bolometric AGN luminosity using a universal conversion factor of 22.4 \citep[median value found in][ and based on a local AGN sample with $L_{X} = 10^{41\--46}~{\rm erg~s^{-1}}$]{Vasudevan2007}, and assuming a ratio of 0.002 between SMBH masses and stellar masses \citep{Marconi2003}. Therefore, we have,

\begin{equation}
\label{eq:eddrat}
\text{\Ledd} = \frac{22.4 \times \text{\Lx}}{10^{38.1}~\text{\ergps} \times 0.002~\frac{\text{\Mstar}}{\text{\Msun}}},
\end{equation}

\noindent where \Ledd\ is the Eddington ratio, and \Lx\ is the intrinsic (absorption-corrected) 2-10 keV X-ray luminosity. We stress that the important parameter we investigate is the specific X-ray luminosity (\Lx/\Mstar), and that the Eddington ratio is solely used for its greater familiarity.

Throughout, we assume a {\it WMAP--7} year cosmology \citep{Larson2011} and a \cite{Salpeter1955} initial mass function (IMF) when calculating galaxy stellar masses and SFRs.

\section{Methodology}
\label{sec:method}

\begin{figure}
  \includegraphics[scale = 0.38]{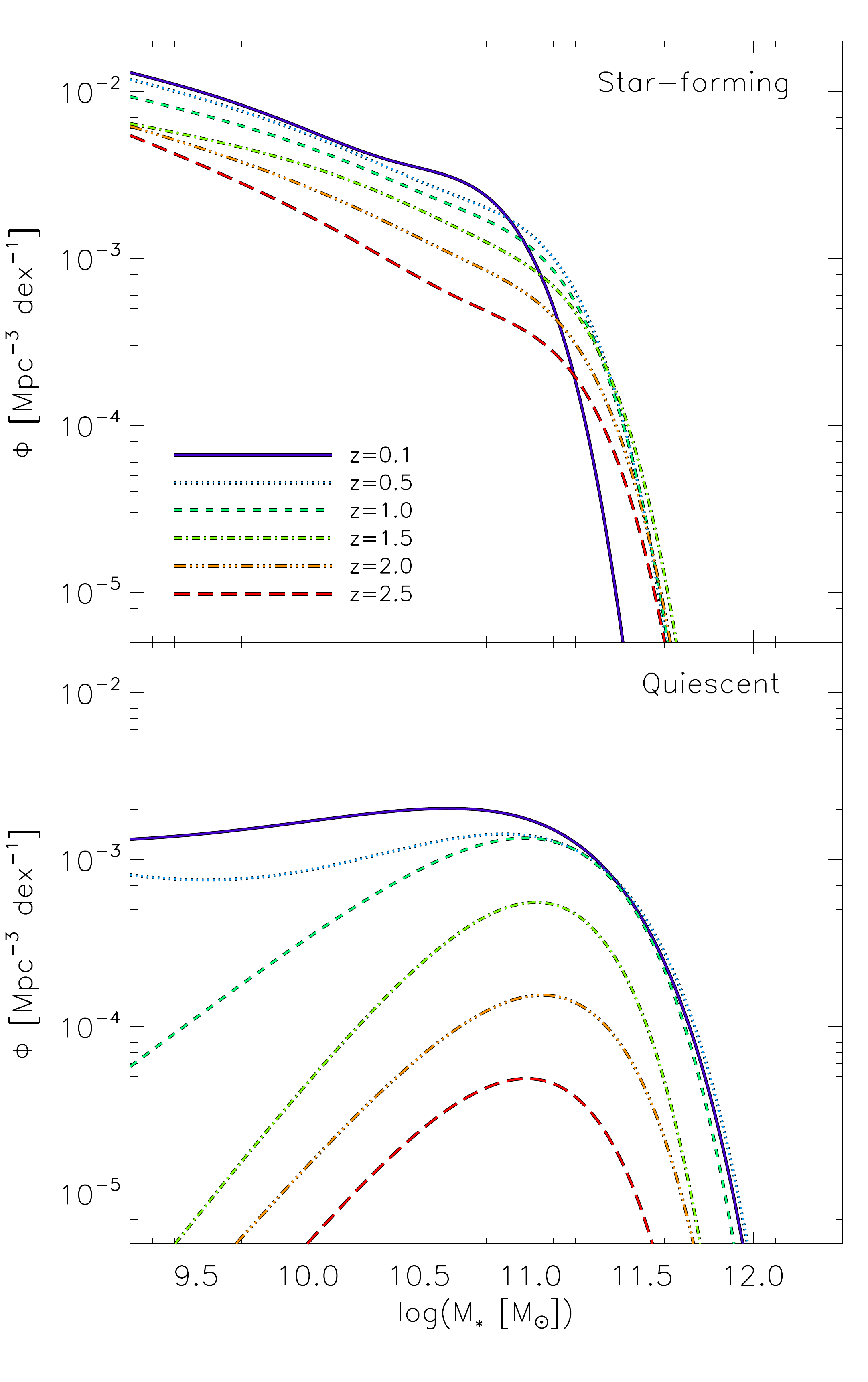}
  \caption{Mass functions employed in our models. {\it Top panel:} mass functions for star-forming galaxies as reported in \protect\cite{Davidzon2017}. Each colour corresponds to a different redshift bin, out to \z=2.5 (see key). {\it Bottom panel:} same as top panel but for the quiescent galaxies.}
  \label{fig:mass_func}
\end{figure}

Many recent studies have used Population Synthesis Models (hereafter PSMs) to investigate the connection between AGNs and host galaxies. PSMs randomly draw properties from probability distribution functions \citep[e.g.][]{Aird2013,Hickox2014,Stanley2015,Jones2017}. Using PSMs we are able to generate populations of $N$ massive mock galaxies, with stellar masses \Mstar, drawn from a mass distribution, for which we randomly allocate Eddington ratios, \Ledd, following an Eddington ratio distribution, and, using Eq.\,\ref{eq:eddrat}, derive X-ray luminosities, \Lx, the histogram of which is proportional to the model X-ray luminosity function. The key point is to optimise the Eddington ratio distribution until the model X-ray luminosity function matches the observed one \citep[e.g.][]{Aird2013, Jones2017}. This can be repeated at different redshifts to investigate any redshift dependence of the Eddington ratio distribution. Because it implies generating a new population of galaxies while optimising the Eddington ratio distribution, it can be computationally expensive. Therefore, we will use an analytical approach \citep[similar to those of e.g.][]{Veale2014,Caplar2015} by directly considering the various distributions as probability density functions. In this study, we fit to the X-ray luminosity functions of \cite{Aird2015} and use the stellar mass functions of \cite{Davidzon2017}. The latter are shown in Fig.\,\ref{fig:mass_func} for various redshifts, out to \z=2.5.

In our analytical approach, we first assume a model \Ledd\ distribution which is described by a set of parameters $\theta$ (for example, for a power law $\theta= \{N, \alpha\}$, where $N$ is the normalisation and $\alpha$ the slope). Our aim is to iterate over $\theta$ to identify the solution that best fits the observed X-ray luminosity function. The model X-ray luminosity function is generated by analytically combining the observed mass function with the \Ledd\ distribution given by the set of parameters $\theta$. To achieve this, we treat the X-ray luminosity function as a probability distribution, $p(L_X)$. The probability of observing an AGN of luminosity $L_X=X$ in a galaxy of mass $M_\ast$ is given by\footnote{To a constant factor given in Eq.\,\ref{eq:eddrat}.},

\begin{equation}
  p(L_X=X) \equiv p(\lambda_{\rm Edd} = \frac{X}{M_\ast}),
\end{equation}

\noindent where $p(\lambda_{\rm Ledd} = X/M_\ast)$ is the probability of $\lambda_{\rm Edd}=X/M_\ast$ calculated with the current set of model parameters, $\theta$. We derive the total probability of observing $L_X=X$ in {\it all} galaxies as the sum of these individual probabilities weighted by the galaxy mass function (which we also treat as a probability distribution, $p(M_\ast$)),

\begin{equation}
  p(L_X=X) = \sum_{Y=M_{\rm Min}}^{M_{\rm Max}} p(\lambda_{\rm Edd} = \frac{X}{Y} | M_\ast=Y)\times p(M_\ast=Y),
\end{equation}

\noindent where $p(M_\ast=Y)$ is the probability of \Mstar=$Y$ (or the mass function evaluated at $Y$). The total model luminosity function derived from a given set of parameters, $\theta$, is then simply the combination of these probabilities (i.e. for $L_X=10^{41}\--10^{46}$~\ergps\ in bins of 0.1 dex).

By taking this approach we avoid having to model $N$ galaxies (where $N\sim\mathcal{O}(10^9)$ to generate sufficient numbers of luminous galaxies), and can simply split our mass function into $\mathcal{O}(10^3)$ bins between $M_{\rm Min}=10^8$~\Msun\ to $M_{\rm Max}=10^{14}$~\Msun\ (where $M_{\rm Min}\ {\rm and}\ M_{\rm Max}$ are chosen to cover the full range of stellar masses). Since we adopt an iterative process to identify the best fitting parameters (see later in this subsection), this factor of $\mathcal{O}(10^{6})$ reduction in the number of required calculations {\it per iteration} dramatically reduces the time taken to converge.

To optimise the parameters that define our Eddington ratio distribution, we performed maximum likelihood estimation. Assuming Gaussian distributions for the uncertainties, the log-likelihood of a set of parameters $\theta$ is defined by -0.5$\chi^2$, where $\chi^2$ is the chi-square between the model and the observed X-ray luminosity functions. We used flat proper (i.e. with finite boundaries) prior distributions for each parameter that defines our model, and checked the posterior distributions to ensure that they are not constrained in any way by the limits of our prior distributions. The parameter space (defined by the flat prior distributions) was explored using a Monte Carlo Markov Chain (MCMC) fully implemented in a {\sc python} application programming interface, {\sc emcee}\footnote{{\sc emcee} is available on-line at \url{http://dan.iel.fm/emcee/current/}} \citep[][]{Foreman-Mackey2013}, that uses the affine invariant MCMC ensemble sampler of \cite{Goodman2010}. We ran our MCMC code using 100 walkers for 10000 steps as a burn-in phase to find the global solution (i.e. these 10000 steps are then discarded), and for 5000 steps as a production phase (i.e. these 5000 steps are kept for analysis). The convergence of each chain was assessed using the \cite{Gelman1992} test. We then extracted the best parameters by fitting a Gaussian function to each sampled posterior distribution, taking the mean and the standard deviation as the best estimate and the 1$\sigma$ uncertainties, respectively, of each parameter. When the posterior distribution of a parameter was flat below or above a given value, we assumed that it is an upper or a lower limit, respectively (again, we ensured that values of these limits are not affected by our prior distributions). This optimisation method was applied to infer the Eddington ratio distribution for both our mass-independent and mass-dependent models.

\subsection{A mass-independent \Ledd\ distribution}
\label{subsec:mass_ind}

As highlighted in the introduction, a number of studies have recently reported a difference in the \Ledd\ distribution of quiescent and star-forming host galaxies \cite[e.g.][]{Georgakakis2014, Wang2016}. We therefore take this ``two-component'' distribution as our starting point (i.e. we do not attempt to model a single universal \Ledd\ distribution for {\it all} galaxies). Most studies use a broken power-law (or related functions) to represent the Eddington ratio distribution of AGNs \citep[e.g.][]{Aird2013, Veale2014, Jones2017, Weigel2017}. As such, we take a conservative approach by assuming a broken power-law for both the star-forming and the quiescent component of our \Ledd\ distribution. Each broken power law is defined as,

\begin{equation}
\label{eq:pl_distrib}
p(\lambda_{\rm Edd}) = \frac{A}{\left(\frac{\lambda_{\rm Edd}}{\lambda_{\rm break}} \right)^{\gamma_{1}}+\left(\frac{\lambda_{\rm Edd}}{\lambda_{\rm break}} \right)^{\gamma_{2}}},
\end{equation}

\noindent where $p(\text{\Ledd})$ is the probability of Eddington ratio, $A$ is the normalisation, $\lambda_{\rm break}$ is the position of the break, and $\gamma_{1}$ and $\gamma_{2}$ are the slopes at low \Ledd\ (i.e. below the break) and high \Ledd\ (i.e. above the break), respectively. As such, each of our broken power law components are represented by four parameters, giving a total of eight free parameters to optimise for this mass-independent model (see \S\,\ref{sec:method} for the optimisation method).

Following empirical results reported in e.g. \cite{Georgakakis2014, Aird2017, Wang2016}, prior to optimisation, we assume that both the normalisation, $A_{\rm SF}$, and the position of the break, $\lambda_{\rm break}^{\rm SF}$, of the star-forming component are each always higher than their quiescent analogues, $A_{\rm Qui}$ and $\lambda_{\rm break}^{\rm Qui}$, respectively. This also has the benefit of reducing the degeneracy of the model. However, it is important to stress that we do not assume any specific parameter values for the ratio between the two normalisations and the two positions of the breaks, we simply incorporate this information in the prior distributions by excluding some parts of the parameter space.

Incorporating the aforementioned assumptions, we optimised the eight free parameters of our mass-independent model. To investigate any redshift evolution of the Eddington ratio distribution, we performed the optimisation at \z=0.3, \z=0.5, \z=0.7, \z=1.0, \z=1.2, \z=1.5, \z=1.7, \z=2.0, \z=2.3, \z=2.5, \z=2.5, \z=2.7, \z=3.0, and \z=3.5, interpolating the mass functions of \cite{Davidzon2017} split between star-forming and quiescent galaxies, and the total X-ray luminosity functions of \cite{Aird2015} at these redshifts. We stress that the Bayesian methodology adopted in \cite{Aird2015} to derive the observed X-ray luminosity functions (see \cite{Aird2015} for details on their method) means that they consider {\it all} X-ray detected sources when deriving their X-ray luminosity functions (which our analysis aims to fit). Therefore, there is no implied mass cut beyond the fact that their sources are detected in X-rays.

\subsection{A mass-dependent \Ledd\ distribution}
\label{subsec:mass_dep}

Motivated by recent studies reporting that the \Ledd\ distribution for star-forming galaxies is mass-dependent \cite[e.g.][]{Bongiorno2016, Aird2017, Georgakakis2017}, we develop a second model that can accommodate three different \Ledd\ distributions in three different mass bins for star-forming galaxies. As with our mass-independent model, we still split in terms of quiescent and star-forming galaxies, but for the star-forming component, we assume instead three different broken power-laws (see Eq.\,\ref{eq:pl_distrib}), one for each of our mass bins, which we define as: low mass (8$<$log(\Mstar/\Msun)$<$10), medium mass (10$<$log(\Mstar/\Msun)$<$11), and high mass (11$<$log(\Mstar/\Msun)$<$12). The choice of the boundaries for our mass bins is arbitrary and independent of any empirical studies, yet aims to cover a large range of stellar masses (i.e. $10^{8}<$\Mstar/\Msun$<10^{12}$).

Since assuming three broken power-laws generates 12 free parameters, we again rely on some assumptions based on the results of \cite{Aird2017} to help prevent degeneracies. First, we require that each broken power-law breaks at different \Ledd\ values, in such a way that a higher mass bin breaks at a lower value of \Ledd\ than its lower mass neighbour. Secondly, we assume that the three Eddington ratio distributions share the same slope at high Eddington ratios. Finally, we assume the normalisation of each of the three broken power-laws is such that the \Ledd\ distribution above the break is always coincident (see Fig.\,\ref{fig:Ledd_mass_dep}). In making these assumptions, we reduce the parameter space to eight free parameters:

\begin{itemize}
\item one normalisation, $A_{\rm SF}$, (from which the others are derived via our third assumption); \\
\item three break positions, $\lambda_{\rm break}^{\rm SF}$, (one for each mass bin and ordered according to our first assumption); \\
\item three power-law slopes below the break, $\gamma_{1}^{\rm SF}$, (which are unconstrained); \\
\item a single shared power-law slope above the break, $\gamma_2^{\rm SF}$.
\end{itemize}

\noindent We stress that all these assumptions are inspired by empirical findings reported by \cite{Aird2017}. We will demonstrate in \S\,\ref{subsubsec:XLF_mass_ind} that our previous model (i.e. the mass-independent model, see \S\,\ref{subsec:mass_ind}) is able to reproduce the observed X-ray luminosity functions out to \z$\sim$2, and recreates both their star-forming and quiescent galaxy components in good agreement with observations, at least out to \z$\sim$1. As such, for this mass-dependent model, we assume that the Eddington ratio distribution for quiescent galaxies is unchanged from our previous model, and only update the star-forming component of the model Eddington ratio distribution by implementing the aforementioned mass-dependency. To do this, we optimise the eight free parameters that describe our mass-dependent Eddington ratio distribution for star-forming galaxies. Again, to investigate any redshift evolution of the mass-dependent Eddington ratio distribution, we performed the optimisation at \z=0.1, \z=0.3, \z=0.5, \z=0.7, \z=1.0, \z=1.3, \z=1.5, and \z=1.7, interpolating the mass functions for star-forming galaxies of \cite{Davidzon2017} and the star-forming component of the X-ray luminosity functions as extracted from our mass-independent model. This second model relies on the ability of our first, mass-independent, model to split the X-ray luminosity functions in terms of star-forming and quiescent galaxies. As we show in \S\,\ref{subsubsec:XLF_mass_ind}, our mass-independent model is unable to reproduce the total X-ray luminosity function beyond \z$\sim$2, and thus cannot be used to reliably split the luminosity function between star-forming and quiescent galaxies. As such, we cannot extend our mass-dependent model beyond \z$\sim$2.

\subsection{Generating galaxies with AGNs}
\label{subsec:PSM}

As we aim to investigate the connection between AGN accretion and host star-formation activity, we must model the star-forming properties of the host galaxies as well as the Eddington ratios of the AGNs. Since the distributions of specific SFRs (i.e SFR relative to the host stellar mass; sSFR) are now well-defined, we can simply adopt published models to generate our galaxy populations. As such, for this part of the model we use a PSM to attribute SFRs to our host galaxies. We use the PSM outlined in \cite{Bernhard2014}, which generates a population of mock star-forming galaxies using the mass functions of \cite{Ilbert2013} and allocates SFRs using the relationship between stellar masses and SFRs, or Main Sequence \citep[MS; e.g.][for the MS]{Daddi2007, Salim2007, Elbaz2011, Rodighiero2011, Sargent2012, Schreiber2015} reported in \cite{Rodighiero2011}. The SFRs are then randomly scattered around the sSFR distribution of \cite{Sargent2012}. This model successfully reproduces the observed infra-red and ultraviolet luminosity functions up to \z$\sim$6 \citep[see][and references therein]{Bernhard2014}. However, for this work, we slightly update the empirical relationships used in \cite{Bernhard2014}. Briefly, we change the mass functions to the more recent ones reported in \cite{Davidzon2017}, we update the MS sSFR distribution to that reported in \cite{Schreiber2015}, and lastly, add a population of quiescent galaxies (using the quiescent galaxy mass functions of \citealt{Davidzon2017}) with sSFRs at least a factor of ten below that of the MS \citep{Ilbert2013}. We confirm that this updated model still reproduces the infra-red and ultraviolet luminosity functions. X-ray luminosities are then allocated using our Eddington ratio distributions derived in \S\,\ref{sec:method}.

Using this PSM, we generate a population of 33,925,192 galaxies (i.e. 32,939,834 -- or 97~per~cent -- star-forming and 985,358 -- or 3~per~cent -- quiescent galaxies), with stellar masses in the range $10^{8}<$\Mstar/\Msun$<10^{14}$. This corresponds to a 50 square degrees blank-field survey out to \z=3. We use a very high (arguably un-physical) upper limit on our range of stellar masses to rule-out the possibility of being affected by any boundary effects. The large relative number of star-forming compared to quiescent galaxies in our model is related to the differences of their respective mass functions, particularly at masses \Mstar$\lesssim 10^{10}$~\Msun\ and \z$\gtrsim$1. As shown in Fig.\,\ref{fig:mass_func}, at these lower masses and higher redshifts, the mass function of star-forming galaxies steeply rises while that of quiescent galaxies sharply decreases. This leads to a large difference in the absolute number of star-forming galaxies compared to that of quiescent galaxies at lower masses (we have a cut at \Mstar=$10^{8}$~\Msun\ in our model). Above a stellar mass of \Mstar$> 10^{10}$~\Msun, we have a total of 3,309,201 galaxies, among which 2,646,102 (80~per~cent) are star-forming and 663,099 (20~per~cent) are quiescent.

\section{Results}
\label{sec:results}

In this section we first present the results of our mass-independent model, followed by the results of our mass-dependent model. We demonstrate that while our mass-independent model can successfully reproduce the observed X-ray luminosity functions out to \z$\approx$2, it cannot at the same time reproduce the observed flat relationship between SFR and AGN luminosity \citep[e.g.][]{Rosario2012,Stanley2015}. Instead, we show that this can be resolved by introducing a mass dependency in the \Ledd\ distribution of AGNs hosted in star-forming galaxies.

\subsection{The mass-independent model}
\label{subsec:mass_indep_results}

\subsubsection{X-ray luminosity functions}
\label{subsubsec:XLF_mass_ind}

\begin{figure*}
  \includegraphics[scale = 0.7]{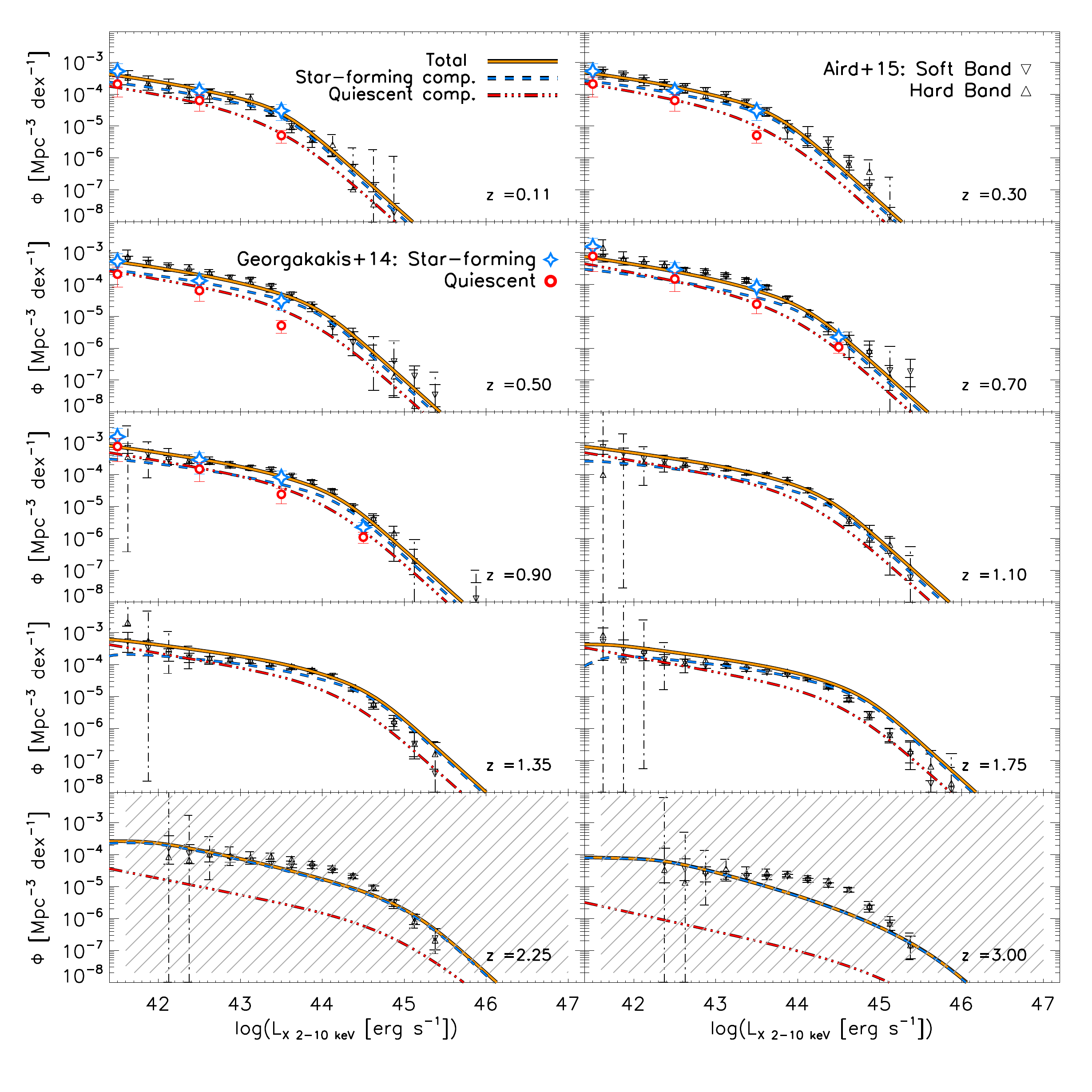}
  \caption{Fit to the X-ray 2-10 keV luminosity functions of \protect\cite{Aird2015} when optimising our mass-independent \Ledd\ distribution model. Each panel represents a different redshift bin out to \z=3, and as indicated at their bottom right-hand side. The downward and upward empty black triangles are the X-ray luminosity functions of \protect\cite{Aird2015} for the soft and the hard band, respectively. The thick orange line in each panel is for the model total X-ray luminosity function given by our mass-independent model, the blue dashed line is for that of the star-forming component and the red triple dashed line shows the quiescent component. The blue empty stars and the red empty circles at \z$<$0.90 are for the star-forming and the quiescent component of the X-ray luminosity functions, respectively, as reported in \protect\cite{Georgakakis2014} out to \z$\sim$1. We hatched the two panels at \z=2.25 and \z=3.0 to indicate that our mass-independent \Ledd\ distribution model is unable to reproduce the observed X-ray luminosity functions at these redshifts (see \S\,\ref{subsubsec:XLF_mass_ind}).}
  \label{fig:XLF_mass_ind}
\end{figure*}

We show in Fig.\,\ref{fig:XLF_mass_ind} the fit to the X-ray luminosity functions for our mass-independent model. Up to \z=1.75 our model X-ray luminosity functions are in very good agreement with those of \cite{Aird2015}. However, beyond that redshift, our mass-independent model is unable to reproduce the observed X-ray luminosity functions. Fig.\,\ref{fig:XLF_mass_ind} also illustrates the model X-ray luminosity functions split between star-forming and quiescent galaxies. Overall, we find that the star-forming galaxies always dominate the X-ray luminosity functions at \Lx$\gtrsim 10^{42.5}$~\ergps, which corresponds to the bulk of the AGN population. The contribution from quiescent galaxies in our mass-independent model, however, is only significant at lower X-ray luminosities (i.e. \Lx$\lesssim 10^{42}$~\ergps).

Beyond \z$\sim$2, our model cannot reproduce the observed X-ray luminosity functions. At the position of the break, our model X-ray luminosity functions at these redshifts are at least a factor of three below the observed ones. We also find that the contribution from quiescent galaxies is very low (by at least a factor of ten) compared to that of the star-forming galaxies. To explain the reasons for this, we examine the stellar mass functions of \citeauthor{Davidzon2017} (\citeyear{Davidzon2017}; see Fig.\,\ref{fig:mass_func}). They report that the contribution from quiescent galaxies to the mass function decreases as the redshift increases (the knee of the mass function for quiescent galaxies is at least a factor of ten in normalisation below that of the star-forming galaxies at \z$\gtrsim$2; see Fig.\,\ref{fig:mass_func}). In our model, this lower contribution from the quiescent galaxies can be compensated by an increase in the normalisation of the model Eddington ratio distribution for quiescent galaxies (as our model relates the mass function to the Eddington ratio distribution). This would increase the contribution from quiescent galaxies to the total model X-ray luminosity functions in order to match the observed ones. However, in our model, the normalisation of the quiescent galaxy component cannot exceed that of the star-forming component (see assumptions in \S\,\ref{subsec:mass_ind}). While looking at the normalisations of both the star-forming ($A_{\rm SF}$) and the quiescent ($A_{\rm Qui}$) component in our model, we find that beyond \z$\sim$2, they are similar (i.e. $A_{\rm SF} \sim A_{\rm Qui}$). As a consequence, the contribution from quiescent galaxies to the total X-ray luminosity functions cannot increase enough to match the observed X-ray luminosity functions. If we relax this assumption, our model finds that the Eddington ratio distribution beyond \z$\sim$2 is fully dominated by quiescent galaxies, with the normalisation of the Eddington ratio distribution for star-forming galaxies to that of quiescent galaxies as low as $\sim$0.01 at \z=2.25 and $\sim$0.001 at \z=2.5. This is at odds with empirical measurements of the Eddington ratio distributions of e.g. \cite{Georgakakis2014} and \cite{Wang2016}, that suggest a more equal splitting between quiescent and star-forming galaxies, as well as infrared studies that find large numbers of high-redshift AGNs residing in star-forming galaxies \citep[e.g.][]{Mullaney2015, Stanley2015, Netzer2016}.

Finally, as a further check, we include in Fig.\,\ref{fig:XLF_mass_ind} the measured X-ray luminosity functions separated into star-forming and quiescent galaxies for \z$<$1 as reported in \cite{Georgakakis2014}. Despite not including this information during our optimisation, we find good agreement between our model and these observed X-ray luminosity functions of star-forming and quiescent galaxies, increasing our confidence in the model at these lower redshifts.

\subsubsection{The Eddington ratio distribution}
\label{subsec:mass_indep_Eddrat}

\begin{figure*}
  \centering
  \includegraphics[scale = 0.45]{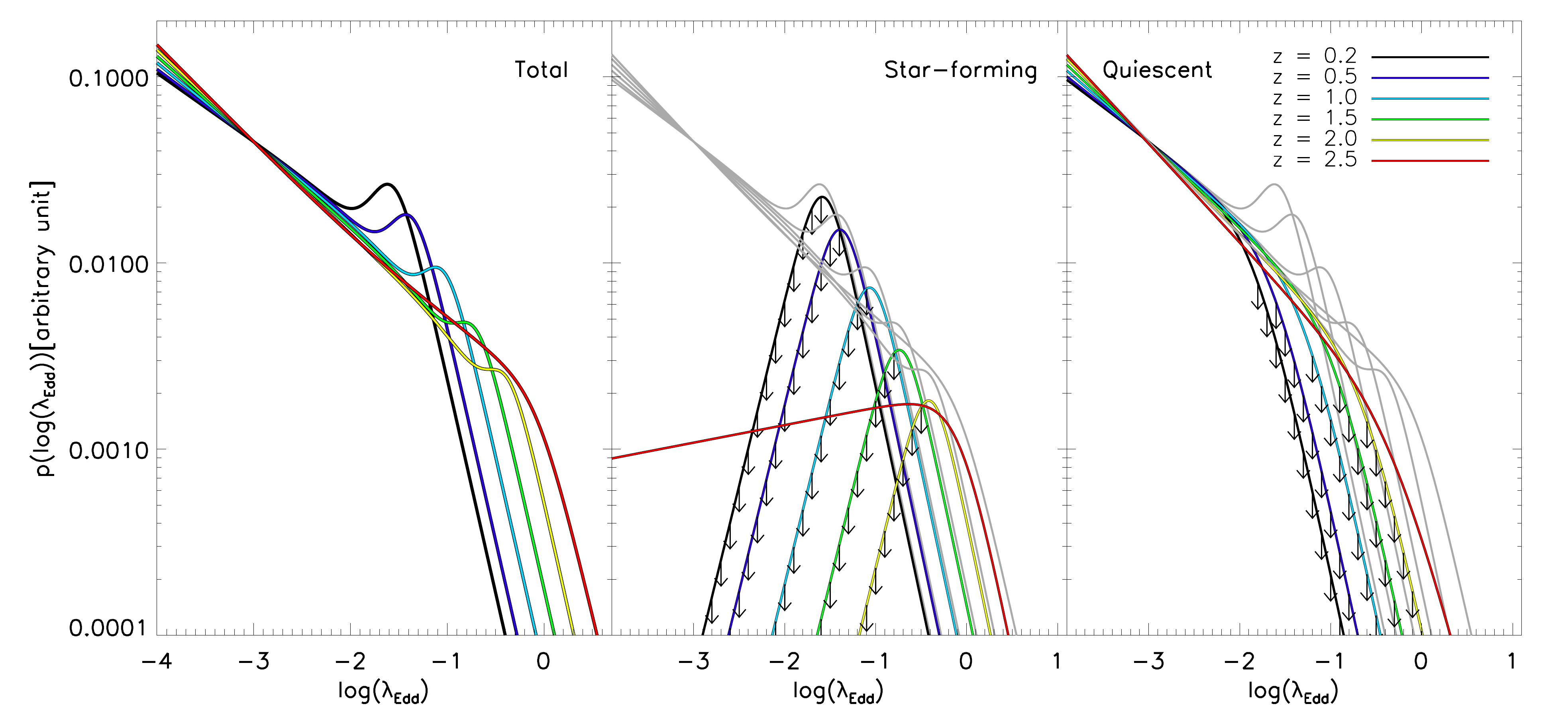}
  \caption{The total mass-independent Eddington ratio distribution (left-hand panel), its star-forming component (central panel), and its quiescent component (right-hand panel) evaluated at \z=0.2, \z=0.5, \z=1.0, \z=1.5, \z=2.0, and \z=2.5 (colour coded; see key). The downward arrows indicate the overall effects of the upper and lower limits found in the parameters that describes these distributions. For the star-forming and quiescent panels, we also indicate with faint grey lines the corresponding total Eddington ratio distribution. The normalisation is such that the total Eddington ratio distribution at \z=0.2 integrates to unity, applying an arbitrary cut at \Ledd=$10^{-7}$. For ease of comparison, we then re-normalise all the distributions at higher redshifts such that their values at log(\Ledd)=-3.0 coincide with that at \z=0.2, hence the arbitrary unit of the ordinate axis.}
  \label{fig:Ledd_mass_ind}
\end{figure*}

We show in \S\,\ref{subsubsec:XLF_mass_ind} that our mass-independent model is able to reproduce the observed X-ray luminosity function out to \z$\sim$2. We now present in Fig.\,\ref{fig:Ledd_mass_ind} the Eddington ratio distributions at various redshifts after optimisation of our mass-independent model (normalised to coincide at \Ledd = $10^{-3}$). The shape of our total Eddington ratio distribution (see left-hand panel in Fig.\,\ref{fig:Ledd_mass_ind}) is consistent with a broken power-law, as requested by our model and in agreement with previous studies \citep[e.g.][]{Aird2013, Jones2017}. However, we find a significant difference in the slope below the break ($\gamma_1$ in Eq.\,\ref{eq:pl_distrib}) of the Eddington ratio distribution for star-forming hosts compared to that of quiescent hosts (see central and right-hand panels in Fig.\,\ref{fig:Ledd_mass_ind} for the Eddington ratio distribution of star-forming and quiescent host galaxies, respectively). We report that below \z$\sim$2, the slope at low-\Ledd\ for the star-forming component is rising (although we are only able to place upper limits). As a consequence, our model Eddington ratio distribution of AGNs in star-forming galaxies is consistent with a peaky\footnote{Hereafter, we refer to the shape of our model Eddington ratio distribution of AGNs hosted by star-forming galaxies as peaky.} distribution, similar to the light-bulb shaped distribution sometimes explored in past studies \citep[e.g.][]{Veale2014, Stanley2015}. This suggests that AGNs hosted in star-forming galaxies have typical \Ledd\ higher than AGNs in quiescent hosts, yet the \Ledd\ of AGNs in quiescent galaxies span a broader range of values. Finally, we notice an overall shift of the knee of the Eddington ratio distribution to higher \Ledd\ values as redshift increases. This implies a typical AGN activity that increases as redshift increases. At \z$\gtrsim$2, we find that the shape of the Eddington ratio distribution of AGNs in star-forming galaxies becomes consistent with a broken power-law, in contrast to the peaky Eddington ratio distribution reported for lower redshifts. However, as demonstrated in \S\,\ref{subsubsec:XLF_mass_ind} our mass-independent model is unable to reproduce the X-ray luminosity functions at these higher redshifts. We report in Table\,\ref{tab:param_mass_ind} the redshift evolution of the parameters that define these Eddington ratio distributions.

\begin{table}
\centering
\caption{Redshift evolution of the parameters that describe the Eddington ratio distribution for star-forming and quiescent galaxies, given by Eq.\,\ref{eq:pl_distrib}. SF and Qui labels stand for star-forming and quiescent, respectively. The slopes and intercepts are given for an evolution in (1+\z). Uncertainties on the parameters are the standard deviation of 1000 Monte Carlo realisations (i.e. 1$\sigma$).}
\label{tab:param_mass_ind}
\begin{tabular}{llll}
\hline
\vspace{0.2cm}
{\sc Parameters}                              & {\sc Intercepts}  & {\sc Slopes}       &        \\
\multirow{2}{*}{log($A_{\rm SF}$)}& \multirow{2}{*}{$-2.07^{\pm0.19}$}  & 0.13$^{\pm0.07}$ & for \z$<$1.7 \\
\vspace{0.2cm}
                                              &   & $-1.15^{\pm0.03}$  & for \z$>$1.7 \\
\multirow{2}{*}{log($A_{\rm Qui}$)}& \multirow{2}{*}{$-2.80^{\pm0.23}$}  & 0.37$^{\pm0.10}$ & for \z$<$1.7 \\
\vspace{0.2cm}
                                              &   & $-1.20^{\pm0.11}$  & for \z$>$1.7 \\
\vspace{0.2cm}
log$(\lambda_{\rm break}^{\rm SF})$   & $-2.36^{\pm0.06}$  & 0.65$^{\pm0.02}$ & \\
\vspace{0.2cm}
log$(\lambda_{\rm break}^{\rm Qui})$   & $-2.62^{\pm0.16}$  & 0.66$^{\pm0.06}$ & \\

\multirow{2}{*}{$\gamma_1^{\rm SF}$} & $\lesssim -3.0$ & 0.0 & for \z$<$1.7 \\
\vspace{0.2cm}
                                              &  $-0.96^{\pm0.51}$ & 0.25$^{\pm0.12}$ & for \z$>$1.7 \\
\vspace{0.2cm}
$\gamma_1^{\rm Qui}$                         & $0.28^{\pm0.05}$ & 0.07$^{\pm0.02}$  &  \\

\multirow{2}{*}{$\gamma_2^{\rm SF}$} & \multirow{2}{*}{$2.29^{\pm0.04}$}  & 0.01$^{\pm0.02}$ & for \z$>$1.7 \\
\vspace{0.2cm}
                                              &   & 0.15$^{\pm0.03}$  & for \z$>$1.7 \\
\multirow{2}{*}{$\gamma_2^{\rm Qui}$}    & $\gtrsim$2.7 & 0.0 & for \z$<$1.7\\
										&	3.75$^{\pm2.33}$  & $-0.53^{\pm0.63}$ & for \z$>$1.7  \\
\hline
\multicolumn{4}{p{7.8cm}}{\footnotesize{{\it Notes:} Slopes and intercepts are given for an evolution as a function of (1+\z). The intercept for \z$>$1.7, when assuming a break in the \z\ evolution of the parameter, is given by the continuity at \z=1.7 (i.e. $[\text{intercept for \z}>1.7] = (1+1.7)\times([\text{slope for \z}<1.7]-[\text{slope for \z}>1.7])+[\text{intercept for \z}<1.7])$.}}
\end{tabular}

\end{table}

\subsubsection{SFR in bins of X-ray luminosities}
\label{subsubsec:SFR_Xray_mass_ind}

\begin{figure*}
  \centering
  \includegraphics[scale = 0.4]{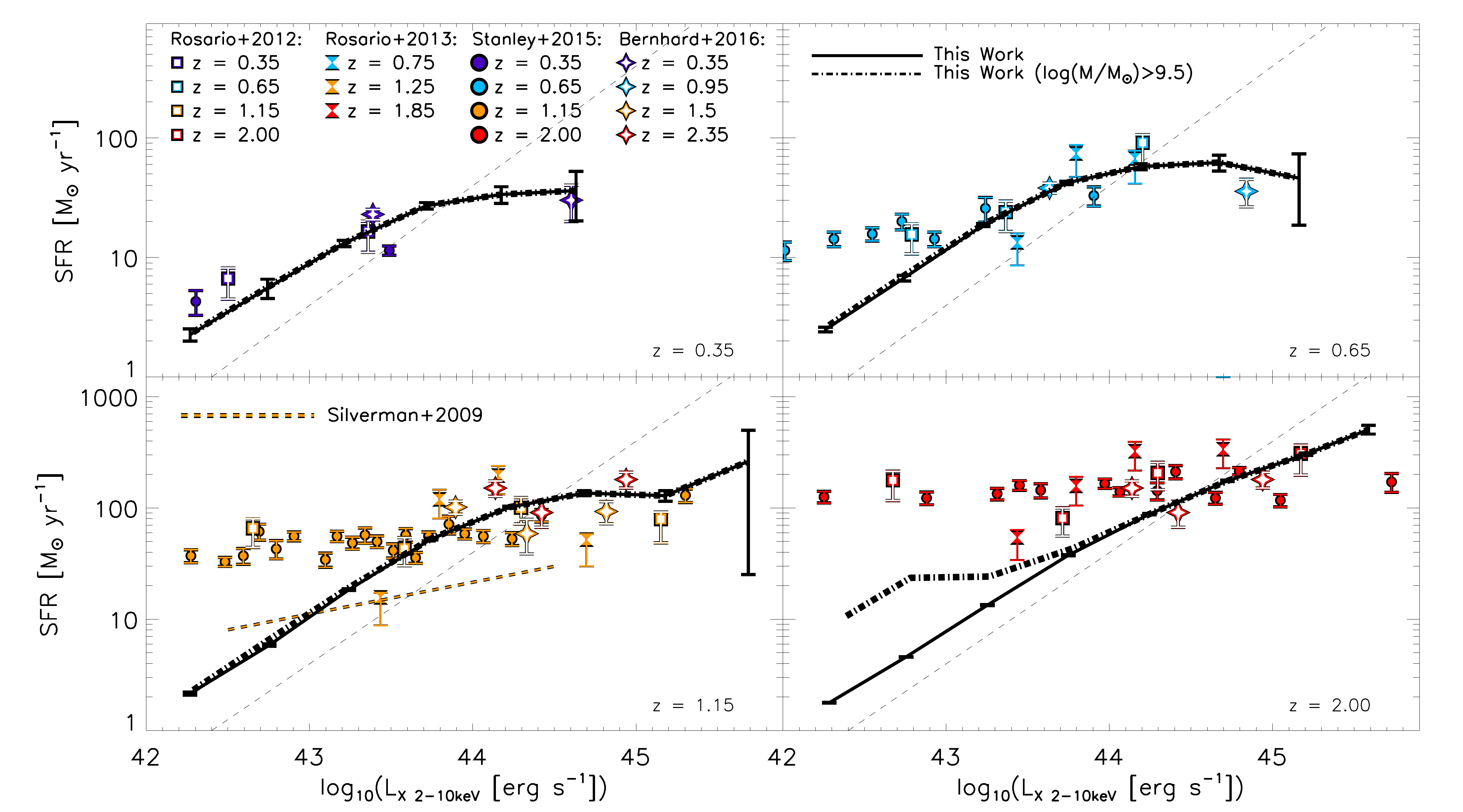}
  \caption{The mean-average SFR in bins of X-ray 2-10 keV luminosity at \z=0.35, \z=0.65, \z=1.15, and \z=2. The black thick lines shows the correlation predicted by our PSM that incorporates AGNs using our mass-independent Eddington ratio distribution split between star-forming and quiescent galaxies. The error bars show the 1$\sigma$ uncertainties on the mean for average SFR. The thick black dashed line is identical than the thick black line but for galaxies with \Mstar$\gtrsim 10^{9.5}$~\Msun. Various symbols are for a compilation of published SFR/X-ray relationships \protect\cite[see keys; i.e.][]{Rosario2012, Rosario2013c, Stanley2015, Bernhard2016}. The orange dashed line at \z=1.15 shows the relationship reported in \protect\cite{Silverman2009}. Finally, the black dashed line represents the correlation displayed by AGN-dominated systems in \protect\cite{Netzer2009}. Our mass-independent model predicts a significant correlation between SFR and X-ray luminosity which is at odds with empirical results for \z$\gtrsim$0.5.}
  \label{fig:SFR_Xray_mass_ind}
\end{figure*}

As previously highlighted, a key test of any model that describes the Eddington ratio distribution is whether it can reproduce other observed features of the AGN population besides just the (X-ray) luminosity functions. Here, we test our PSM (which incorporates pre-defined sSFR distributions and our own optimised mass-independent Eddington ratio distribution; see \S\,\ref{subsec:PSM}) by assessing whether it can reproduce the observed flat relationship between SFR and X-ray luminosity as reported in many observational studies \citep[e.g.][]{Rosario2012,Rosario2013c,Stanley2015,Bernhard2016}. We do this by calculating the mean-average SFRs of the AGNs in our PSM in bins of 0.5~dex in X-ray luminosity, taking into account both star-forming and quiescent galaxies (see \S\,\ref{subsec:PSM}). As shown in Fig.\,\ref{fig:SFR_Xray_mass_ind}, our mass-independent model predicts a strong correlation between these SFRs and X-ray luminosities (with an average slope of $\sim$0.6), at least up to \z$\sim$2. This is in good agreement with empirical results for our lowest redshift bin (i.e. $z=0.35$), but strongly contradicts the flat observed relationship between SFR and X-ray luminosity at higher redshifts \citep[e.g.][]{Rosario2012,Stanley2015}. It is important to note that other studies also find a similar strongly increasing relationship between SFR and X-ray luminosity when employing a peaky (i.e. similar in shape than our peaky distribution; see central panel in Fig.\,\ref{fig:Ledd_mass_ind}) Eddington ratio distribution for star-forming galaxies \citep[e.g.][]{Veale2014, Stanley2015}. We will show in \S\,\ref{sec:discussion} that this strong correlation is due to the AGN host mass distribution and the SFR/stellar-mass relationship (i.e. MS) induced by the peaky Eddington ratio distribution found for our mass-independent model.

As SFR is correlated to stellar mass via the MS, it is possible that the empirical flat relationship between SFR and X-ray luminosity is a consequence of an observational bias toward higher mass galaxies. To test this, we recalculate from our model the average SFR in bins of X-ray luminosity, but now excluding galaxies with \Mstar$< 10^{9.5}$~\Msun. This mass limit roughly corresponds to that reached in the highest redshift bins of \cite{Stanley2015}. We show in Fig.\,\ref{fig:SFR_Xray_mass_ind} that although this mass cut causes the model SFR/X-ray relationship to rise slightly at \z=2.0 and \Lx$\lesssim 10^{43.5}$~\ergps, it is not sufficient to reproduce the observed flat relationship between SFR and X-ray luminosity.

\subsection{The mass-dependent model}
\label{subsec:mass_dep_results}

Since our mass-independent model fails to reproduce the X-ray luminosity functions at \z$\gtrsim$2 and is unable to reproduce the flat relationship between SFR and X-ray luminosity (at \z$>$0.5), we consider whether relaxing the mass-independence requirement (as suggested by recent studies, e.g. \citealt{Bongiorno2016, Aird2017, Georgakakis2017}) can help to resolve these issues. As explained in \S\,\ref{subsec:mass_dep}, for this model we assume that the quiescent component remains unchanged from the mass-independent model meaning that we only fit the star-forming X-ray luminosity functions at \z$<$2. These are extracted from the results of our mass-independent model, and is justified by the ability of our mass-independent model to reproduce the X-ray luminosity functions of star-forming and quiescent galaxies to \z$\sim$1 from \cite{Georgakakis2014}.

\subsubsection{X-ray luminosity functions}
\label{subsubsec:XLF_mass_dep}

\begin{figure*}
  \centering
  \includegraphics[scale = 0.7]{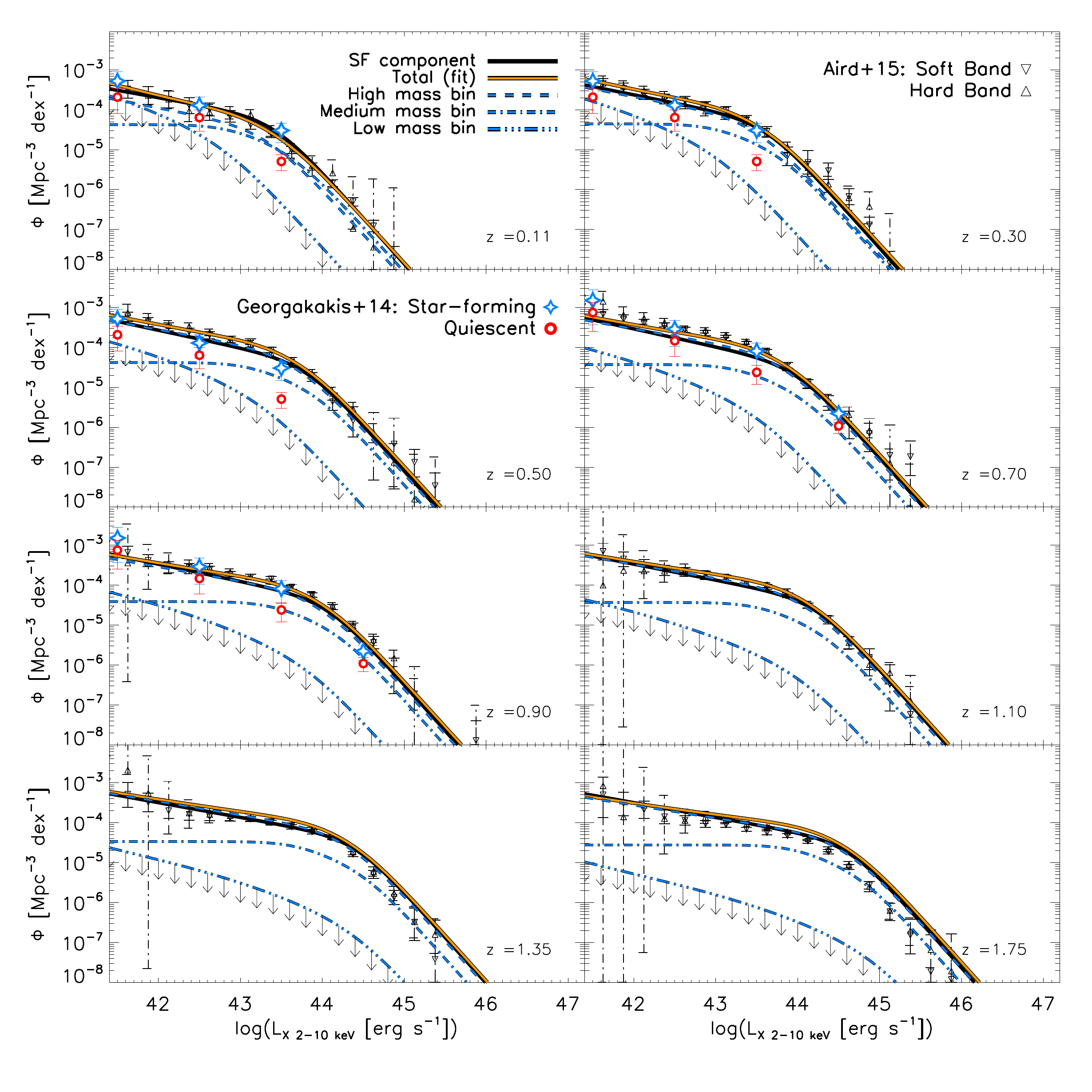}
  \caption{Fit to the star-forming component of the X-ray 2-10 keV luminosity functions using our mass-dependent Eddington ratio distribution model up to \z=1.75. Each panel shows a different redshift bin as indicated on their bottom right-hand side. The black line in each panel shows the star-forming component of the X-ray luminosity function (derived from our mass-independent model), and the orange line shows the best fit given by our mass-dependent model. The dashed, single-dot dashed, and triple-dot dashed lines show the X-ray luminosity functions for our highest (11$<$log(\Mstar/\Msun)$<$12), medium (10$<$log(\Mstar/\Msun)$<$11), and lowest (8$<$log(\Mstar/\Msun)$<$10) mass bin, respectively. The black arrows indicate the overall effects of upper and lower limits on the parameters that define our mass-dependent Eddington ratio distribution. The blue stars and the red circles show the X-ray luminosity functions for the star-forming and quiescent galaxies, respectively, as reported in \protect\cite{Georgakakis2014}. The light grey downward and upward triangles show the total X-ray luminosity functions of \protect\cite{Aird2015} for the soft and the hard band, respectively.}
  \label{fig:XLF_mass_dep}
\end{figure*}

We show in Fig.\,\ref{fig:XLF_mass_dep} the fit to the star-forming component of the X-ray luminosity functions using our mass-dependent Eddington ratio distribution model to \z=1.75. Again, we do not extend our mass-dependent model beyond \z$\sim$2 since it relies on the ability of our previous -- mass-independent -- model to split the X-ray luminosity functions in terms of star-forming and quiescent components (see \S\,\ref{subsec:mass_dep}). Out to \z$\sim$2, our mass-dependent model does a good job at fitting the star-forming X-ray luminosity functions in all our redshift bins. Since our mass-dependent model assumes an \Ledd\ distribution split into three different mass bins (see \S\,\ref{subsec:mass_dep}), we are also able to derive the X-ray luminosity function of each mass bin (see Fig.\,\ref{fig:XLF_mass_dep}). In doing so, our model predicts that the highest mass galaxies (i.e. \Mstar$>10^{11}$~\Msun) dominate the star-forming X-ray luminosity functions across almost the full range of X-ray luminosities (i.e. $10^{41}<$\Lx$<10^{46}$~\ergps). The exception being in our lowest redshift bin (i.e. \z=0.11) where $10^{10}<$\Mstar/\Msun$<10^{11}$ galaxies contribute marginally more than our highest mass galaxies at \Lx$\gtrsim10^{43.5}$~\ergps. In general, however, these medium mass galaxies only contribute significantly (i.e. $>10$~per~cent) at luminosities above the knee of the X-ray luminosity functions (i.e. \Lx$>10^{44}$~\ergps). By contrast, the lowest mass galaxies (i.e. \Mstar$<10^{10}$~\Msun) provide almost no contribution to the X-ray luminosity functions of star-forming galaxies at \Lx$\gtrsim 10^{42.5}$\ergps\ (i.e. $<10$~per~cent), although we are only able to place upper limits on this contribution from lower mass galaxies.

Our mass-dependent model suggests that AGNs hosted by star-forming galaxies with \Mstar$>10^{11}$~\Msun\ dominate the X-ray luminosity functions. This is consistent with \cite{Georgakakis2017}, which demonstrates that AGNs with \Lx$> 10^{41\--42}$~\ergps\ significantly contribute to the AGN host stellar mass function for masses \Mstar$> 10^{11}$~\Msun. This is also consistent with results from \cite{Bongiorno2016} which show that the host stellar mass function of AGNs with \Lx$> 10^{43}$~\ergps\ peaks at \Mstar$\sim 10^{11}$~\Msun, out to \z=2. However, the vast majority of AGN host galaxies with \Lx$< 10^{43}$~\ergps\ display typical stellar masses in the range \Mstar$\sim 10^{10\--11}$~\Msun\ \citep[e.g.][]{Bongiorno2016, Aird2017a}, which contrasts with our findings of a model X-ray luminosity function dominated by host galaxies with \Mstar$>10^{11}$~\Msun. However, our model is limited by the choice for the boundaries of our mass bins. Therefore, should we change our highest mass bin to incorporate galaxies with \Mstar$\sim 10^{10.5}$~\Msun, we would expect to find that galaxies with \Mstar$> 10^{10.5}$~\Msun\ dominate the X-ray luminosity functions, in agreement with recent observational studies \citep[e.g.][]{Bongiorno2016, Aird2017a, Wang2016}. 

\subsubsection{SFR in bins of X-ray luminosities}
\label{subsubsec:SFR_Xray_mass_dep}

\begin{figure*}
  \centering
  \includegraphics[scale = 0.4]{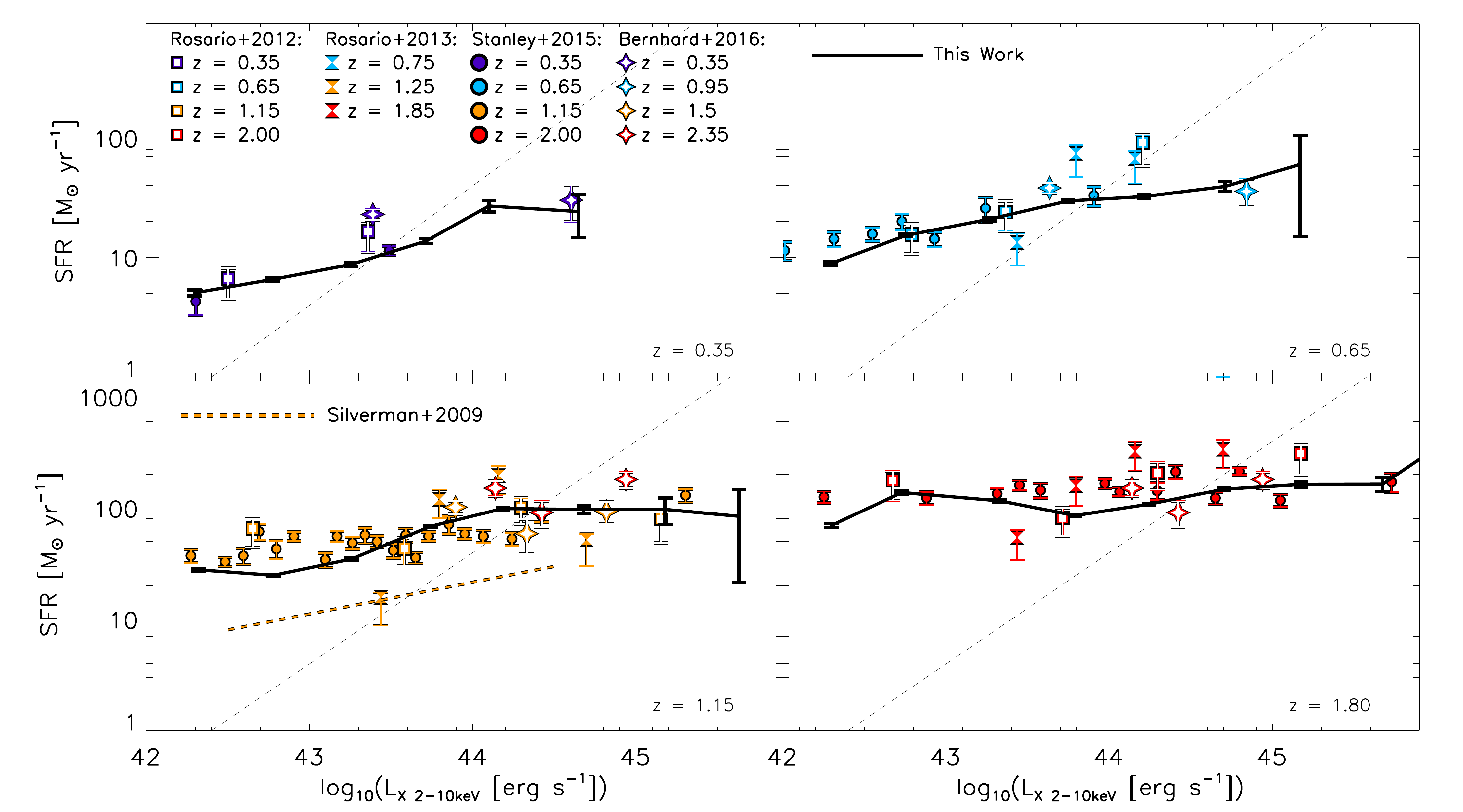}
  \caption{The mean-average SFR in bins of X-ray 2-10 keV luminosities predicted by our mass-dependent model (black thick line) at \z=0.35, \z=0.65, \z=1.15, and \z=2.0. The error bars represent the 1$\sigma$ uncertainties on the mean for average SFR. The various symbols are from a compilation of published relationships \protect\cite[see keys; i.e.][]{Rosario2012, Rosario2013c, Stanley2015, Bernhard2016}. The orange dashed line shows the relationship between average SFR and X-ray luminosity found in \protect\cite{Silverman2009}, and the black dashed line represents the relationship for AGN-dominated systems, as reported in \protect\cite{Netzer2009}. Although it is not implemented in the model, our mass-independent model successfully reproduces the observed flat SFR/X-ray relationship out to \z=2.}
  \label{fig:SFR_Xray_mass_dep}
\end{figure*}

Having confirmed that our mass-dependent model is able to reproduce the X-ray luminosity functions for star-forming galaxies up to \z$\sim$2, we now consider whether the corresponding PSM reproduces the observed flat relationship between SFR and X-ray luminosity \citep[e.g.][]{Rosario2012, Stanley2015}. We show in Fig.\,\ref{fig:SFR_Xray_mass_dep} the mean-average SFR of AGNs split into 0.5~dex-wide bins of X-ray luminosity at similar redshifts as those of \cite{Stanley2015}. Again, we include quiescent galaxies in our model when calculating these averages. Contrary to our mass-independent model, we find that our mass-dependent model predicts a flat relationship between SFR and X-ray luminosity in very good agreement with observations (see Fig.\,\ref{fig:SFR_Xray_mass_dep}). It is important to stress that the model was {\it not} optimised to recreate the flat SFR/X-ray luminosity relationship. Therefore, we conclude that the mass-dependent model is able to reproduce the X-ray luminosity functions for star-forming galaxies (with a good agreement with observations at least up to \z$\sim$1) while also {\it independently} reproducing the observed flat relationship between SFR and X-ray luminosity out to \z$\sim$2.

\subsubsection{The evolution of the \Ledd\ distribution}

\begin{figure*}
  \centering
  \includegraphics[scale = 0.5]{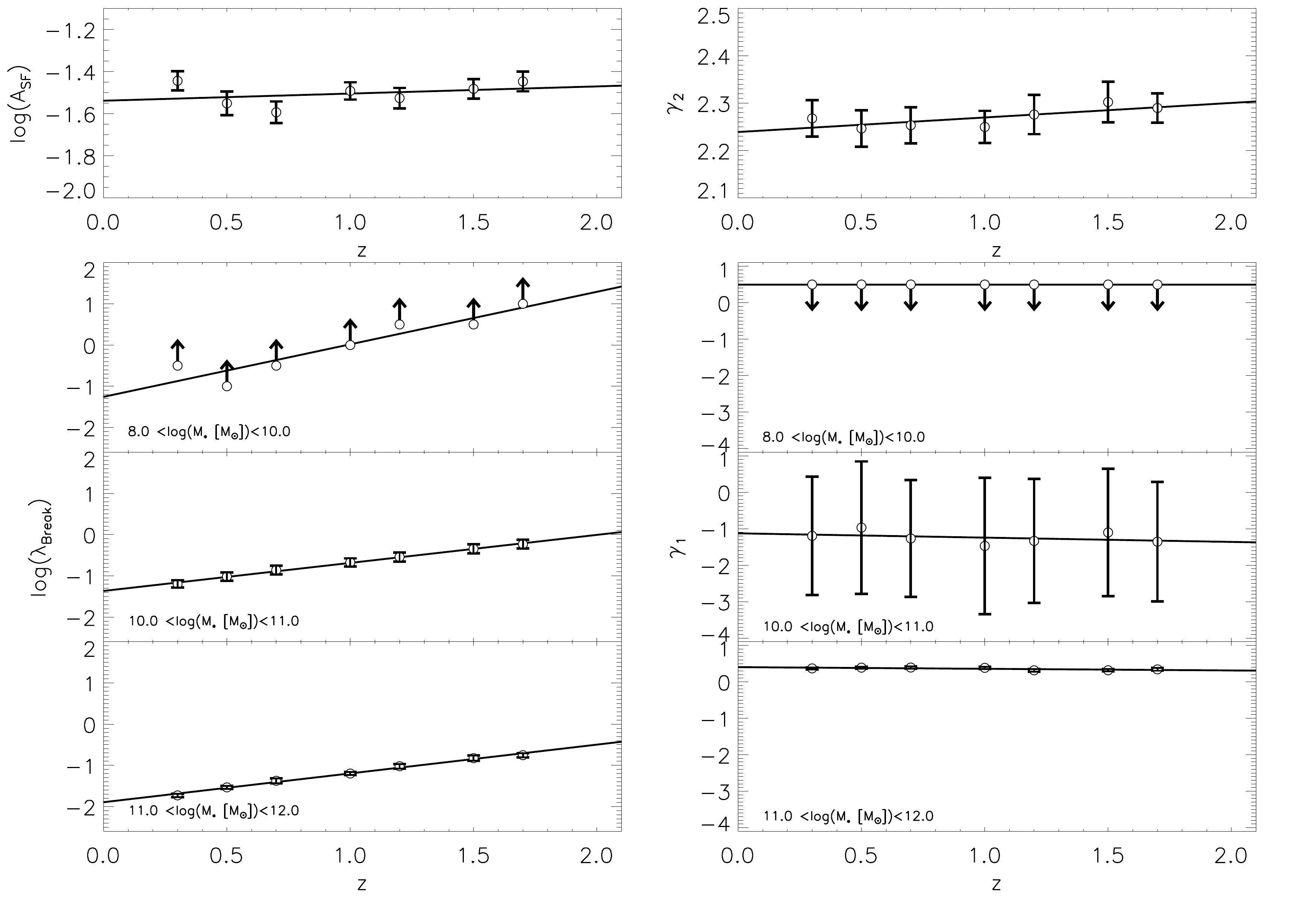}
  \caption{Redshift evolution of the parameters that define our mass-dependent Eddington ratio distribution for star-forming galaxies at \z=0.1, \z=0.3, \z=0.5, \z=0.7, \z=1.0, \z=1.3, \z=1.5, and \z=1.7. Error bars show the 3$\sigma$ uncertainties. The upward and downward arrows indicate the lower and upper limits, respectively. The two top panels present the shared parameters among our three mass bins, i.e. the normalisation, $\log_{A_{\rm SF}}$, and the slope at high Eddington ratio, $\gamma_2^{\rm SF}$. The left-hand panels are the three different break positions $\lambda_{\rm break}^{\rm SF}$, one for each mass bin (as indicated in the bottom left-hand side in each panel), and the right-hand panels are the same but for that of the lower Eddington ratio slope, $\gamma_1^{\rm SF}$. The black lines show the best fit with redshift of each parameter.}
  \label{fig:zevol_mass_dep}
\end{figure*}

\begin{figure*}
  \centering
  \includegraphics[scale = 0.45]{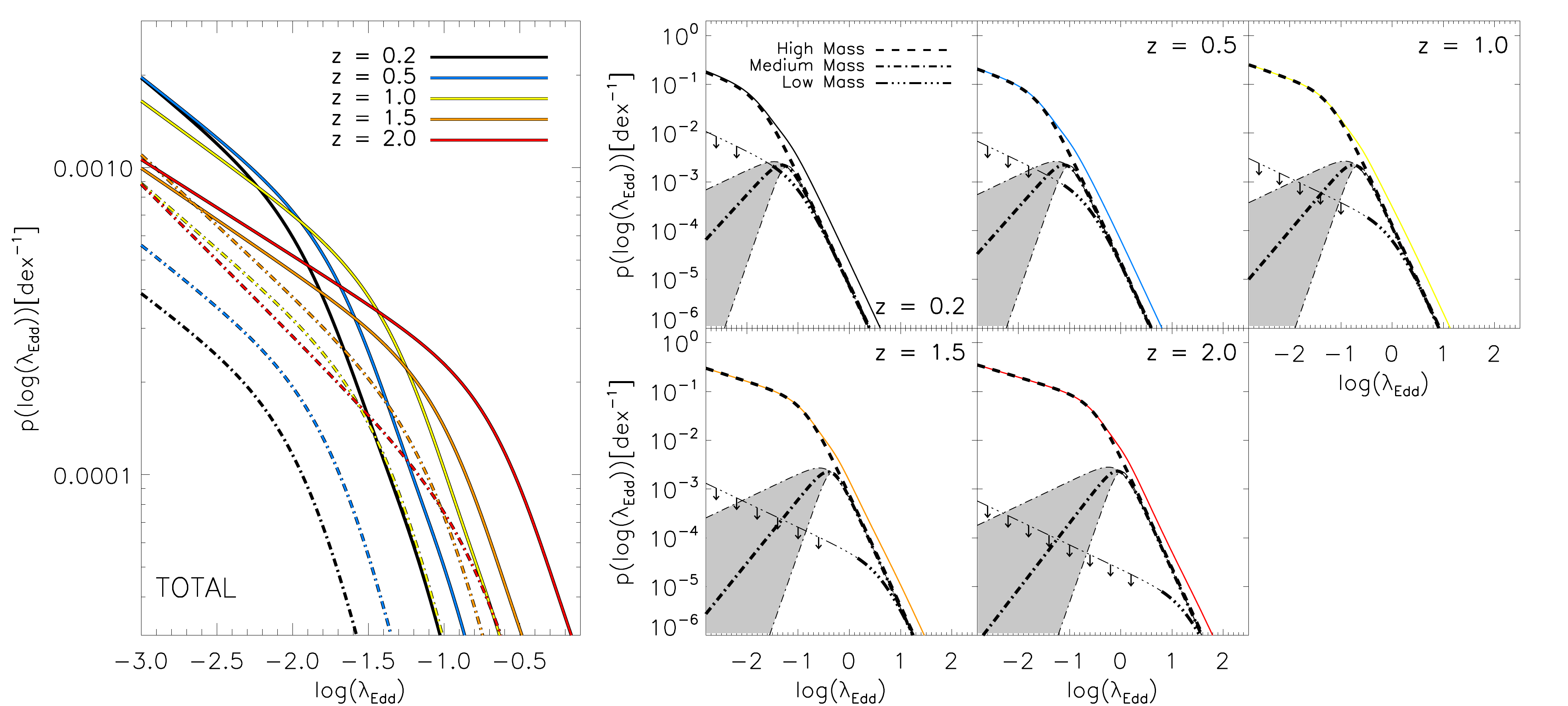}
  \caption{The Eddington ratio probability distributions of star-forming and quiescent galaxies in our mass-dependent model at \z=0.2, \z=0.5, \z=1.0, \z=1.5, and \z=2.0. The left-hand panel shows both the star-forming (plain lines) and the quiescent components (single-dashed lines) of the Eddington ratio probability distribution. The normalisation is such that the total Eddington ratio distribution at each redshift integrates to unity after applying an arbitrary cut at \Ledd = $10^{-7}$. The right-hand panels are, for each redshift (indicated on their top right-hand corner), the contribution of the different mass bins to the Eddington ratio distributions of AGNs hosted in star-forming galaxies. In each of these panels, the dashed line is for the highest mass bin (11$<$log(\Mstar/\Msun)$<$12), the single-dotted dashed line is for the medium mass bin (10$<$log(\Mstar/\Msun)$<$11), and the triple-dotted dashed line is for the lowest mass bin (8$<$log(\Mstar/\Msun)$<$10). The downward arrows for the slope at low-\Ledd\ in our lower mass bin indicate that we have upper limits. The grey area illustrates the large uncertainties for the slope at low-\Ledd\ in our medium mass bin (see Fig.\,\ref{fig:zevol_mass_dep}). We normalised each of the {\bf total} Eddington ratio distribution such that it integrates to unity, applying an arbitrary cut at \Ledd=$10^{-7}$.}
  \label{fig:Ledd_mass_dep}
\end{figure*}

Having demonstrated that our mass-dependent model is able to reproduce both the X-ray luminosity functions for star-forming galaxies and the flat SFR/X-ray luminosity relationship to $z\sim2$ we now explore how the eight parameters (see \S\,\ref{subsec:mass_dep}) that define our mass-dependent model change with redshift. We show in Fig.\,\ref{fig:zevol_mass_dep} the redshift evolution of these parameters. We report that the overall normalisation of the star-forming component ($A_{\rm SF}$) increases very slightly with increasing redshift. The position of the break ($\lambda_{\rm break}^{\rm SF}$) in each mass bin shifts toward higher \Ledd\ values. This is consistent with the overall increase of the position of the break observed in our mass-independent model (i.e. all mass bins collapsed together; see \S\,\ref{subsec:mass_indep_Eddrat}), suggesting a higher typical accretion rate for AGNs at higher redshifts. Regarding the slopes at lower \Ledd, ($\gamma_1^{\rm SF}$), we find that it does not evolve with redshift for any of our mass bins. However, although $\gamma_1^{\rm SF}$ for our highest mass bin is well constrained, we have large uncertainties associated with $\gamma_1^{\rm SF}$ for our medium mass bin, and upper limits for our lowest mass bin (see Fig.\,\ref{fig:zevol_mass_dep}). These weak constraints on $\gamma_1^{\rm SF}$ for our medium and lowest mass bins are a direct consequence of their low contributions to the total Eddington ratio distribution for star-forming galaxies (see Fig.\,\ref{fig:Ledd_mass_dep}), as was hinted by their low contribution to the total X-ray luminosity functions (see \S\,\ref{subsec:mass_dep_results}). We also find that the shared slope at high \Ledd\ ($\gamma_2^{\rm SF}$) does not change significantly with redshift. Overall, our findings suggest that there is a suppression of low-\Ledd\ (i.e. \Ledd$\lesssim$0.1) for AGNs hosted in lower mass galaxies (i.e. \Mstar$\lesssim 10^{10\--11}$~\Msun). Finally, we performed a linear fit of the redshift evolution of each parameter, using the standard deviation of 1000 Monte Carlo realisations for the uncertainties on the fitting parameters. We report the results of this fit in Table\,\ref{tab:param_mass_dep}.

Each of the aforementioned trends can be seen in the evolution of the overall Eddington ratio distribution, which we plot out to \z=2 in Fig.\,\ref{fig:Ledd_mass_dep}. In this figure, we also plot the Eddington ratio distributions of quiescent host galaxies found in our mass-independent model at similar redshifts. In Fig.\,\ref{fig:Ledd_mass_dep}, for the left-hand panel, each distribution is normalise such that the total Eddington ratio distribution (i.e. the combination of the star-forming and quiescent component) integrates to unity after applying an arbitrary cut at \Ledd = $10^{-7}$. For the right-hand panels, each {\bf total} distribution is also normalised such that it integrates to unity after applying an arbitrary cut at \Ledd = $10^{-7}$.

\begin{table}
\centering
\caption{Redshift evolution of the parameters that describe the Eddington ratio distribution of star-forming galaxies for our mass-dependent model. The slopes and intercepts are given for an evolution as a function of (1+\z). Uncertainties on the fitting parameters are extracted via 1000 Monte Carlo realisations (i.e. 1$\sigma$).}
\label{tab:param_mass_dep}
\begin{tabular}{llll}
\hline
\vspace{0.2cm}
{\sc Parameters }          & {\sc Intercepts} & {\sc Slopes }      \\

\vspace{0.2cm}
log($A_{\rm SF}$) & -1.57$^{\pm0.08}$ & 0.03$^{\pm0.04}$ \\
\vspace{0.2cm}
log($\lambda_{\rm break}^{\rm low~mass}$)    & $>$-2.53 & 1.27           \\
\vspace{0.2cm}
log($\lambda_{\rm break}^{\rm medium~mass}$) & -2.05$^{\pm0.15}$ & 0.68$^{\pm0.08}$                       \\
\vspace{0.2cm}
log($\lambda_{\rm break}^{\rm high~mass}$)   & -2.60$^{\pm0.08}$ & 0.70$^{\pm0.04}$              \\
\vspace{0.2cm}
$\gamma_1^{\rm low~mass}$                        & $<$0.5 & 0.0             \\
\vspace{0.2cm}
$\gamma_1^{\rm medium~mass}$                      & -1.0$^{\pm2.67}$ & -0.12$^{\pm1.30}$             \\
\vspace{0.2cm}
$\gamma_1^{\rm high~mass}$                       & 0.45$^{\pm0.05}$ & -0.04$^{\pm0.02}$       \\
\vspace{0.2cm}
$\gamma_2$ & 2.21$^{\pm0.06}$ & -0.03$^{\pm0.03}$            \\

\hline
\multicolumn{4}{p{7.8cm}}{\footnotesize{{\it Notes:} Slopes and intercepts are given for an evolution as a function of (1+\z). The intercept for \z$>$1.7 when assuming a break in the \z\ evolution of the parameters is given by the continuity at \z=1.7 (i.e. $[\text{intercept for \z}>1.7] = (1+1.7)\times([\text{slope for \z}<1.7]-[\text{slope for \z}>1.7])+[\text{intercept for \z}<1.7])$.}}
\end{tabular}
\end{table}

\section{Discussion}
\label{sec:discussion}

\subsection{Further checks to our mass-dependent model}

\begin{figure}
  \centering
  \includegraphics[scale=0.43]{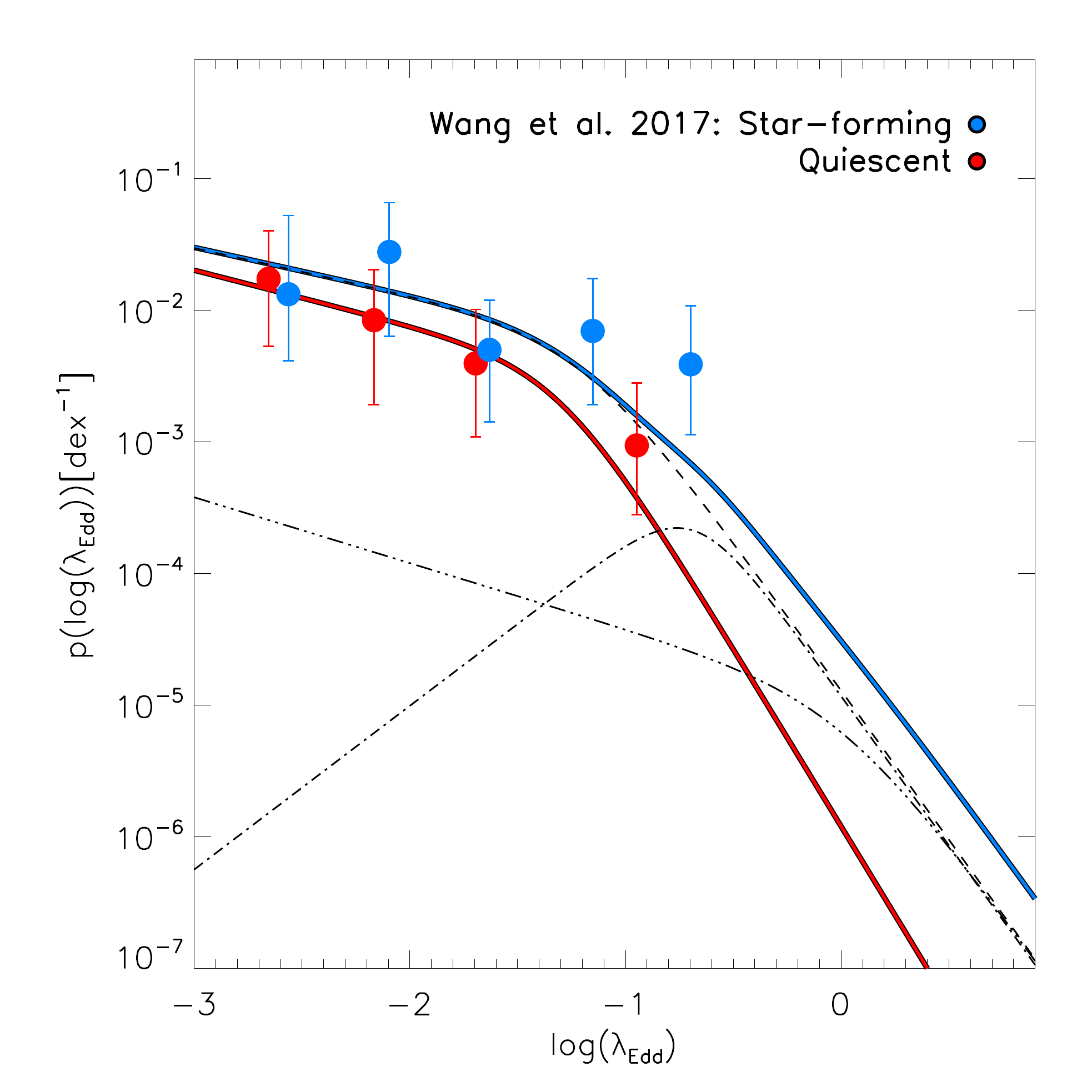}
  \caption{Comparison of our mass-dependent model Eddington ratio distribution to empirical results of \protect\cite{Wang2016} at \z=1. The blue thick line is the total model Eddington ratio distribution for star-forming galaxies, while the dashed line, the dotted-dashed line, and the triple-dotted dashed line indicate our highest, medium, and lowest mass bin contributions, respectively. The blue circles are empirical data of the star-forming component from \protect\cite{Wang2016} at \z$\sim$1, while the red circles are that of the quiescent component.}
  \label{fig:comp_wang}
\end{figure}

As a further check of our mass-dependent model, we compared the model Eddington ratio distribution at \z=1 to the empirical one reported in \cite{Wang2016}. As our various assumptions for this model are based on observations from \cite{Aird2017}, comparing our mass-dependent Eddington ratio distribution to that empirical of \cite{Wang2016} -- instead of \cite{Aird2017} -- constitutes a more independent test for our model. We show this comparison in Fig.\,\ref{fig:comp_wang}. We find very good agreement between our model Eddington ratio distributions to that of \cite{Wang2016} at \z=1 for both star-forming and quiescent component (at least for log(\Ledd)$<$-0.5). This strongly supports our mass-dependent model for the Eddington ratio distribution of star-forming galaxies (since the mass-independent model was predicting a peaky Eddington ratio distribution for star-forming galaxies which would contrasts with empirical results from \citealt{Wang2016}). We predict that the low-\Ledd\ end of the Eddington ratio distribution for star-forming galaxies is dominated by galaxies with \Mstar$\gtrsim 10^{11}$~\Msun. However, \cite{Wang2016} reported that their sample of AGNs have typical stellar masses between $10^{10.5}<$\Mstar/\Msun$<10^{11}$ (once corrected for the differences in the IMFs between \citealt{Salpeter1955} for this work and \citealt{Chabrier2003} for \citealt{Wang2016}), which is half a dex below our highest mass bin. As mentioned in \S\,\ref{subsubsec:XLF_mass_dep}, this discrepancy in the host stellar masses between our model and observations is a consequence of the chosen boundaries for each of our mass bins.

\begin{figure*}
  \centering
  \includegraphics[scale = 0.4]{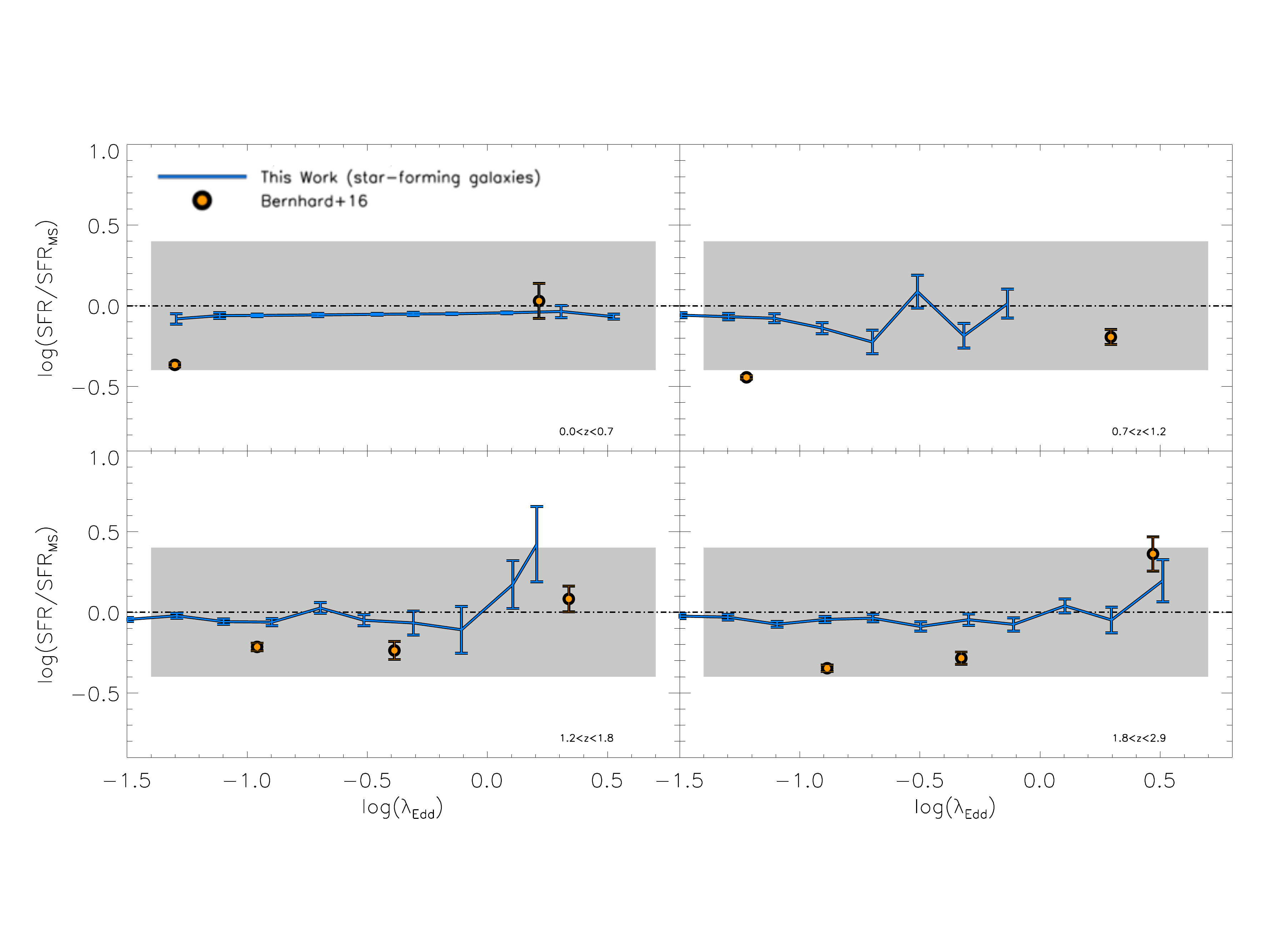}
  \caption{The normalised average SFR versus the Eddington ratio up to \z$\sim$3. The blue lines show the prediction for our mass dependent model. the orange circles show the results report in \protect\cite{Bernhard2016}. The grey area represent the scatter around the MS as reported in \protect\cite{Schreiber2015}, the dashed black line being the MS.}
  \label{fig:SFR_Eddrat_mass_dep}
\end{figure*}

As a final check of our mass-dependent model, we show in Fig.\,\ref{fig:SFR_Eddrat_mass_dep} how the normalised average SFR (i.e. SFR relative to that of the MS) changes with \Ledd\ in our mass-dependent model. Although it is not incorporated in the optimisation, we predict a slight enhancement of average normalise SFR at higher \Ledd\ (i.e. \Ledd$\gtrsim$1) compared to that of their lower \Ledd\ counterpart (at least at \z$\gtrsim$1.2). This slight enhancement of normalised average SFR at higher \Ledd\ is also observed in \cite{Bernhard2016}. In Fig.\,\ref{fig:SFR_Eddrat_mass_dep} we only consider star-forming galaxies for our model since we do not have a good prescription for the normalised SFRs of quiescent galaxies. This is a possible reason for the discrepancy at lower \Ledd\ between our model and the empirical results (i.e. that does contain quiescent galaxies). We also stress that in Fig.\,\ref{fig:SFR_Eddrat_mass_dep} our highest redshift bin, i.e. 1.8$<$\z$<$2.9 constitutes an extrapolation of our model. However, the results are still consistent with observations (i.e. a slight enhancement of the normalised average SFR at higher \Ledd).

\subsection{Why our mass-independent model predicts a strong SFR/X-ray relationship}

One important result of our mass-independent model is that the Eddington ratio distribution for star-forming galaxies is better represented by a peaky distribution (i.e. a narrow distribution), rather than the more common broken power-law (i.e. a broader distribution). To explain this, we refer to \citeauthor{Caplar2015} (\citeyear{Caplar2015}; see also \citealt{Weigel2017}), where they demonstrate that the steepness of the faint-end slope of the X-ray luminosity function is determined by either the low-mass end of the mass function or the low-\Ledd\ slope of the Eddington ratio distribution (since combined together to form the X-ray luminosity function) whichever is the steeper. Since that the faint-end of the observed X-ray luminosity function of \cite{Aird2015} flattens at higher redshifts whilst, in contrast, the low-mass end of the mass function for star-forming galaxies steepens with redshift (see Fig.\,\ref{fig:mass_func}), our mass-independent model attempts to reduce the contribution from star-forming galaxies at lower \Ledd, leading the peaky Eddington ratio distribution for star-forming galaxies found by our mass-independent model. However, beyond \z$\sim$2, the low-mass end of the mass function for star-forming galaxies of \cite{Davidzon2017} is already too steep compared to the flat faint-end of the X-ray luminosity function, such that our model is unable to reproduce the observed X-ray luminosity function. The steepening of the mass functions combined with the flattening of the X-ray luminosity functions suggests a mass-dependent Eddington ratio distribution for star-forming galaxies.

\begin{figure*}
  \centering
  \includegraphics[scale = 0.4]{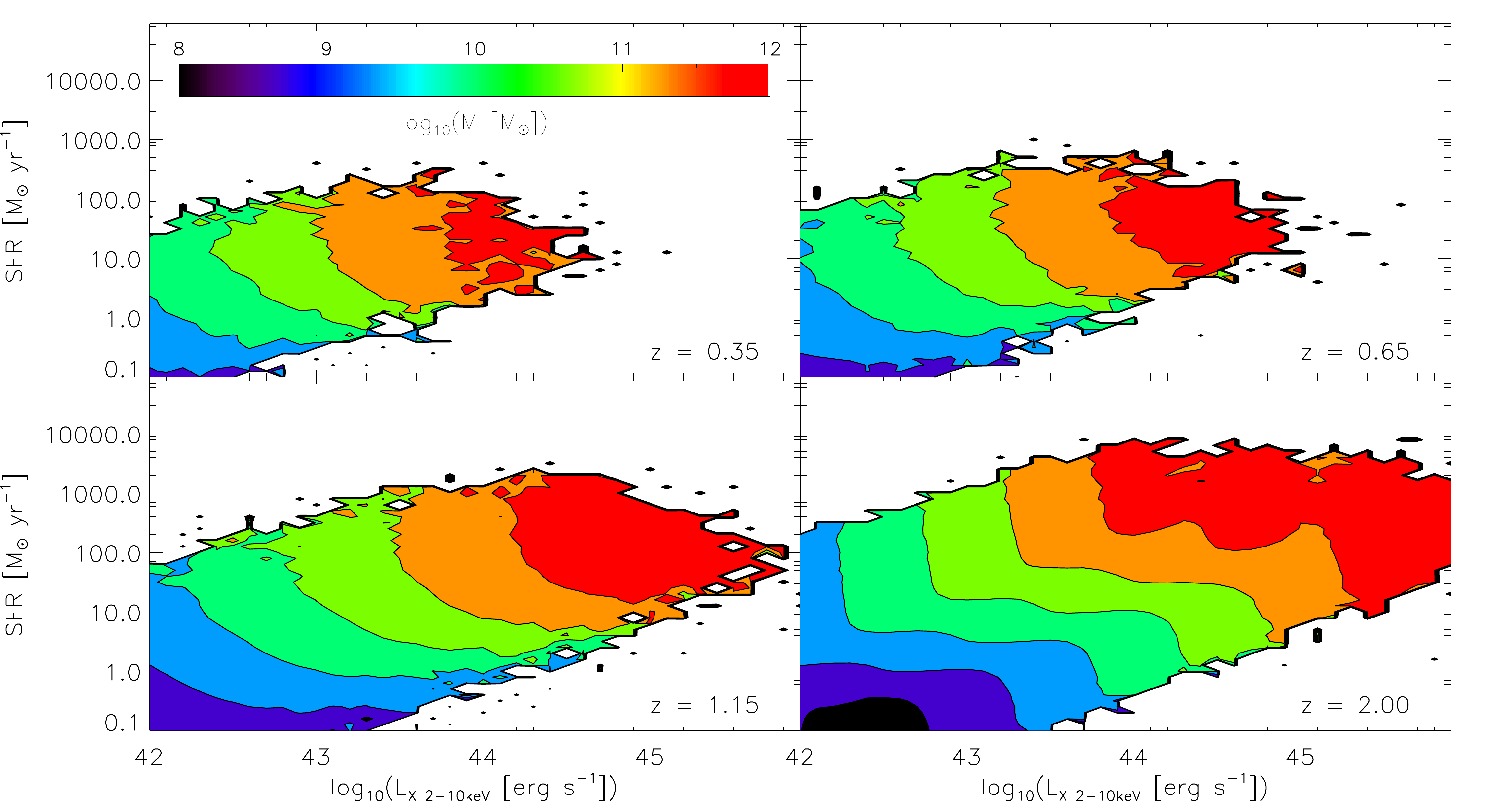}
  \caption{Stellar mass distributions (colour coded) in the SFR/X-ray luminosity parameter space, using our PSM and assuming our mass-independent model for the Eddington ratio distribution. Each panel shows a different redshift, as indicated on their bottom-right hand side. Each discrete colour indicates a stellar mass density in bins of log(\Mstar)=0.5 (see colour-bar).}
  \label{fig:SFR_Xray_mass_distrib}
\end{figure*}

We further saw that our mass-independent model failed to reproduce the observed flat SFR/X-ray relationship (see Fig.\,\ref{fig:SFR_Xray_mass_ind}). To explain why this happens, we show in Fig.\,\ref{fig:SFR_Xray_mass_distrib} which galaxy masses populate different parts of the SFR/X-ray plane. This plot shows that the mass-independent model produces a strong correlation between X-ray luminosity and stellar mass. This is a direct result of the narrower peaky (in contrast to a broader broken power law) Eddington ratio distribution for star-forming galaxies that this model demands in order to reproduce the observed X-ray luminosity functions, at least out to \z$\sim$2. The narrowness of this Eddington ratio distribution means that a galaxy of a given mass can only produce an AGN within a limited range of luminosities. When we then include the correlation between SFR and stellar mass for MS galaxies \citep[e.g.][]{Schreiber2015}, the consequence is a correlation between SFR and X-ray luminosity. This correlation between SFR and X-ray luminosity was also noted by \cite{Veale2014} when using their light-bulb Eddington ratio model, which is similar in shape to our peaky distribution for star-forming host galaxies.

Overall, there is a large discrepancy with our mass-independent model that demands a narrow distribution for the Eddington ratio distribution of star-forming galaxies in order to reproduce the X-ray luminosity functions, but also requires a broader Eddington ratio distribution to reproduce the flat SFR/X-ray relationship. This strongly suggests that the Eddington ratio distribution {\it must be} somehow mass-dependent, such that low Eddington ratios are suppressed in low mass galaxies to be able to reproduce the X-ray luminosity functions while still being broad to reproduce the flat relationship between SFR and X-ray luminosity. Our mass-dependent model provides further constrains on how the Eddington ratio distribution for star-forming galaxies should be mass-dependent.

\subsection{Caveats to our models}

While deriving the \Ledd\ distribution, our model is limited by the functional form chosen for each component of the \Ledd\ distribution and the associated assumptions used to avoid degeneracies during fitting (see \S\,\ref{sec:method}). An obvious caveat to our model is that we cannot explore the full range of possible Eddington ratio distributions. However, we find that, should we relax the assumptions made prior to the fit of the X-ray luminosity functions, our solutions are too degenerate to form a set of meaningful parameters for the Eddington ratio distributions, meaning that the X-ray luminosity functions lack the information needed to constrain the Eddington ratio distribution and its mass-dependency. Using these assumptions is therefore a requirement in our modelling approach. As stressed in \S\,\ref{sec:method}, all our assumptions are motivated by the findings of recent studies \citep[e.g.][]{Jones2017,Aird2017}, and the results predicted by the mass-dependent model are in good agreement with many recent observational studies, i.e. the observed X-ray luminosity functions out to \z$\sim$2 -- including when split in terms of star-forming and quiescent galaxies \citep[e.g.][]{Georgakakis2014,Aird2015}, the flat relationship between SFR and X-ray luminosity \citep[e.g.][]{Rosario2012,Stanley2015}, the empirical Eddington ratio distribution split between star-forming and quiescent component at \z=1 \citep{Wang2016}, and the enhancement of the star-forming properties at higher \Ledd\ \citep{Bernhard2016}. As such while we do not claim that our Eddington ratio distributions are universal, they provide a simple means by which to explore the AGN-galaxy connection.

One of the main caveats for our mass-dependent model is that we are limited by the choice of the stellar mass bins. As a consequence, we find that our dominant component for the Eddington ratio distribution and the X-ray luminosity function is our highest mass bin with \Mstar$>10^{11}$~\Msun. This is qualitatively consistent with observations \citep[e.g.][]{Georgakakis2014, Aird2015, Wang2016}, yet roughly half a dex above the observed typical host mass for AGNs. Investigating this in more detail would require additional mass bins. However, this would also generates a larger number of free parameters, and therefore degeneracies. We have chosen these mass bins to be broad enough to cover a wide range of masses, yet probing any mass dependency of the Eddington ratio distribution for AGNs in star-forming hosts.

\subsection{Extending our mass-dependent model to higher redshifts}
\label{subsec:extension}

We recall that we fit the mass-dependent model to the X-ray luminosity functions of star-forming galaxies only, which was isolated using our mass-independent model. The results of our mass-dependent model therefore depends on the reliability of our mass-independent model to separate the X-ray luminosity functions into star-forming and quiescent components (see \S\,\ref{sec:method}). Thankfully, we can verify this up to \z$\sim$1 using the results reported in \cite{Georgakakis2014}. However, beyond \z$\sim$1 we do not have the empirical results to perform this check, while at \z$\gtrsim$2 our mass-independent model is actually inconsistent with observations (i.e. too many AGNs in quiescent galaxies). Faced with this situation where we cannot reliably exploit the results of our mass-independent model, we now investigate whether we can fit the observed X-ray luminosity functions at \z=2.25 using both a mass-dependent star-forming component {\it and} a mass-independent quiescent component \Ledd\ distributions. This results in a total of 12 free parameters to optimise (i.e. eight parameters for the star-forming component and four parameters for the quiescent one). We use the same assumptions for the \Ledd\ distribution of star-forming galaxies as described in \S\,\ref{sec:method}, and demand that the normalisation of the Eddington ratio distribution for quiescent galaxies lies below that of star-forming galaxies (i.e. consistent with the lower redshift bins).

\begin{figure}
  \centering
  \includegraphics[scale = 0.3]{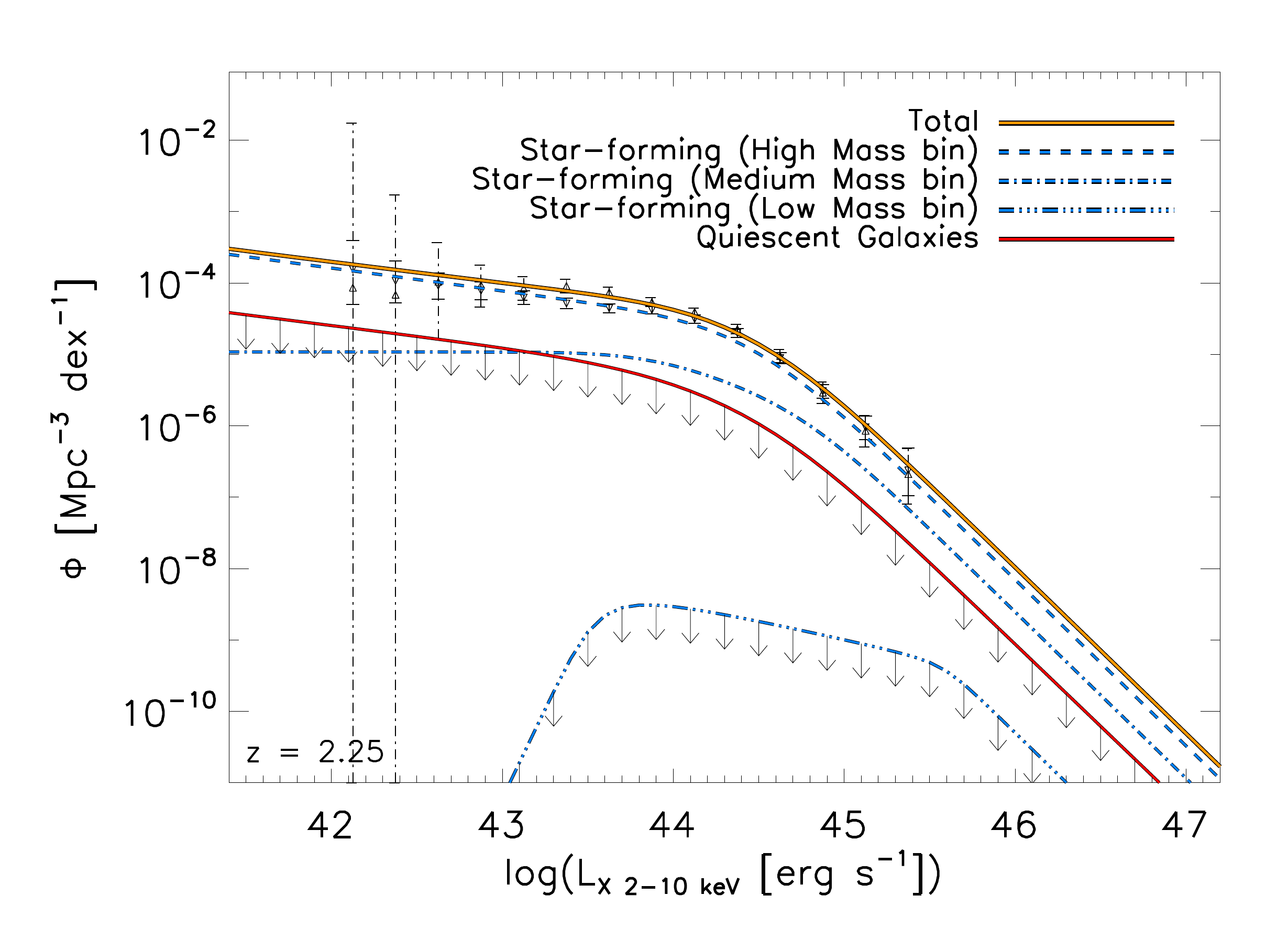}
  \caption{Fit of the X-ray luminosity functions at \z=2.25 of \protect\cite{Aird2015} for our model assuming a mass-dependent Eddington ratio distribution for star-forming galaxies and a mass-independent one for that of quiescent galaxies. The black downward and upward triangles show the X-ray luminosity functions of \protect\cite{Aird2015} for the soft and the hard band, respectively. The orange line shows the total X-ray luminosity functions derived using our model, the blue lines are for that of the star-forming component in each different mass bin (see keys), and the quiescent component is shown with a red line. Downward black arrows indicate upper limits.}
  \label{fig:XLF_fullfit}
\end{figure}

We show in Fig.\,\ref{fig:XLF_fullfit} the results of this new model. Contrary to our mass-independent model, we are now able to reproduce the total observed X-ray luminosity functions at \z=2.25 by placing the majority of AGNs in star-forming galaxies. In particular, we find that the X-ray luminosity functions at \z=2.25 is dominated by the highest mass star-forming galaxies, with a smaller contribution from medium mass star-forming galaxies. The contribution from our lowest mass bin is consistent with zero. Similarly, we can only place upper limits on the contribution from quiescent galaxies to the total X-ray luminosity functions at \z=2.25. We conclude, therefore, that while we are able to reproduce the total X-ray luminosity functions using a mass-dependent \Ledd\ distribution for star-forming galaxies (and ensuring that it is dominated by star-forming galaxies), this solution is degenerate with the level of contribution from quiescent galaxies. To break this degeneracy will require the separation of the high-redshift X-ray luminosity functions into quiescent and star-forming components (i.e. as performed by \cite{Georgakakis2014} at \z$<$1).

\section{Conclusion}
\label{sec:conclusion}

Motivated by recent results reporting a different Eddington ratio distribution for star-forming and quiescent galaxies \citep[][]{Wang2016,Aird2017}, we attempt to constrain these distributions by using an analytical model to fit the observed X-ray luminosity functions of \cite{Aird2015}, assuming the galaxy mass functions of \cite{Davidzon2017}. Our first model assumes two mass-independent \Ledd\ distributions: one for star-forming galaxies and another for quiescent galaxies. After optimisation, we find that this model is able to reproduce the X-ray luminosity functions out to \z$\sim$2 (see \S\,\ref{subsubsec:XLF_mass_ind}), but demands a peaky distribution for the Eddington ratio distribution of star-forming galaxies. Despite this, our mass-independent model fails to reproduce the observed flat SFR/X-ray relationship when incorporated into a PSM for galaxies (see \S\,\ref{subsubsec:SFR_Xray_mass_ind}). We argue that this failure arises because a mass-independent model places too many low-luminosity AGNs in low mass galaxies (see \S\,\ref{sec:discussion}). To resolve this problem, we develop a second model in which the Eddington ratio distribution for star-forming galaxies is allowed to be different in three mass bins (motivated by observation from e.g. \citealt{Aird2017, Georgakakis2017}). By suppressing low Eddington ratio AGNs in low mass galaxies, this mass-dependent model is able to reproduce the X-ray luminosity functions for star-forming galaxies while simultaneously reproducing the observed flat relationship between SFR and X-ray luminosity (see \S\,\ref{subsubsec:SFR_Xray_mass_dep}). Finally, we also find that the mass-dependent model is consistent with empirical Eddington ratio distribution for star-forming and quiescent galaxies at \z=1, and is able to reproduce the slight enhancement of normalised average SFR at higher \Ledd, as reported in \cite{Bernhard2016}.

Overall, we conclude that, in our model, {\it a suppression of lower AGN accretion activity in low mass galaxies} is required to reproduce both the observed X-ray luminosity functions and the observed flat SFR/X-ray relationship.

\section{acknowledgements}

We thank the referee for the useful comments that help to significantly improve the quality of the paper. EB thanks C. Tadhunter and D. Alexander for the helpful discussions. EB thanks the University of Sheffield for the support grant.

\bibliographystyle{mnras}

\bibliography{./biblio}

\end{document}